\documentclass[aps,prd,superscriptaddress,twocolumn,nofootinbib]{revtex4-1}
\usepackage[utf8]{inputenc}
\usepackage{dsfont}
\usepackage{amsfonts}
\usepackage{amsmath}
\allowdisplaybreaks[4]        
\usepackage{amssymb}
\usepackage{euscript}     
\usepackage{braket}
\usepackage{starfont}
\usepackage{color,soul}         
\usepackage{tensor}        
\usepackage{amsthm}
\usepackage{enumitem}
\usepackage{graphicx}
\usepackage{slashed}
\usepackage{leftidx}
\usepackage{subfigure}
\usepackage{bbm}
\definecolor{outerspace}{rgb}{0.25, 0.29, 0.3}
\definecolor{scarlet}{rgb}{1.0, 0.13, 0.0}
\usepackage[header,title,page,titletoc]{appendix}  
\definecolor{princetonorange}{rgb}{1.0, 0.56, 0.0}
\definecolor{WildStrawberry}{rgb}{1.0, 0.26, 0.64}
\definecolor{rossocorsa}{rgb}{0.83, 0.0, 0.0}
\definecolor{navyblue}{rgb}{0.0, 0.0, 0.5}
\usepackage{float}
\usepackage[paper=letterpaper,margin=1in]{geometry}
\usepackage{mathtools}

\newcommand{\fig}[1]{Fig.~\ref{#1}}
\newcommand{\tab}[1]{Table~\ref{#1}}

\DeclareMathAlphabet{\pazocal}{OMS}{zplm}{m}{n}

\newcommand{\bea}{\begin{eqnarray}}

\newcommand{\eea}{\end{eqnarray}}
\newcommand{\ba}{\begin{eqnarray}}
\newcommand{\ea}{\end{eqnarray}}

\newcommand{\be}{\begin{equation}}
\newcommand{\ee}{\end{equation} }
\newcommand{\beqa}{\begin{eqnarray}}
\newcommand{\eeqa}{\end{eqnarray}}
\newcommand{\beqar}{\begin{eqnarray*}}
\newcommand{\eeqar}{\end{eqnarray*}}

\newcommand{\mx}{\mathrm{x}}

\newcommand{\veff}{V_\text{eff.}}
\newcommand{\vt}{\vartheta}

\makeatletter
\newcommand{\dal}{\mathop{\mathpalette\dal@\relax}}
\newcommand{\dal@}[2]{%
  \begingroup
  \sbox\z@{$\m@th#1\square$}%
  \dimen0=\fontdimen8
    \ifx#1\displaystyle\textfont\else
    \ifx#1\textstyle\textfont\else
    \ifx#1\scriptstyle\scriptfont\else
    \scriptscriptfont\fi\fi\fi3
  \makebox[\wd\z@]{%
    \hbox to \ht\z@{%
      \vrule width \dimen0
      \kern-\dimen0
      \vbox to \ht\z@{
        \hrule height \dimen0 width \ht\z@
        \vss
        \hrule height 2\dimen0
      }%
      \kern-2.5\dimen0
      \vrule width 2.5\dimen0
    }%
  }%
  \endgroup
}
\makeatother

\usepackage{hyperref}
\hypersetup{
    colorlinks,
    citecolor=blue,
    filecolor=navyblue,
    linkcolor=navyblue,
    urlcolor=navyblue
}

\begin{document}

\title
{Accuracy of the slow-rotation approximation for black holes in modified gravity in light of astrophysical observables}
\author{Pablo A. Cano}
\email{pabloantonio.cano@kuleuven.be}
\affiliation{Instituut voor Theoretische Fysica, KU Leuven. Celestijnenlaan 200D, B-3001 Leuven, Belgium. }

\author{Alexander Deich}
\email{adeich2@illinois.edu}
\affiliation{Illinois Center for Advanced Studies of the Universe, Department of Physics, University of Illinois at Urbana-Champaign, Urbana, IL 61801, USA}

\author{Nicol\'as Yunes}

\affiliation{Illinois Center for Advanced Studies of the Universe, Department of Physics, University of Illinois at Urbana-Champaign, Urbana, IL 61801, USA}


\begin{abstract}
Near-future, space-based, radio- and gravitational-wave interferometry missions will enable us to rigorously test whether the Kerr solution of general relativity accurately describes astrophysical black holes, or if it requires some kind of modification. At the same time, recent work has greatly improved our understanding of theories of gravity that modify the Einstein-Hilbert action with terms quadratic in the curvature, allowing us to calculate black hole solutions to (essentially) arbitrary order in a slow-rotation expansion. 
Observational constraints of such quadratic gravity theories require the calculation of observables that are robust against the expansion order of the black hole solution used.  

We carry out such a study here and determine the accuracy with respect to expansion order of ten observables associated with the spacetime outside a rotating black hole in two quadratic theories of gravity, dynamical-Chern-Simons and scalar-Gauss-Bonnet gravity. We find that for all but the most rapidly rotating black holes, only about the first eight terms in the spin expansion are necessary to achieve an accuracy that is better than the statistical uncertainties of current and future missions. 
\end{abstract}
\maketitle

\section{Introduction}\label{sec:intro}
From gravitational wave detectors, such as LIGO/VIRGO~\cite{Barausse:2020rsu,danzmann_2016}, to very long baseline interferometers, like the Event Horizon Telescope~\cite{https://doi.org/10.48550/arxiv.2212.05118}, we have never been able to better probe the extreme gravity environment near black holes (BHs)~\cite{will_2014,Johnson_2020}.  The remarkable precision of these, as well as future space-based detectors, allow us to interrogate Einstein's theory of general relativity (GR) to a finer degree than ever before~\cite{will_2014,https://doi.org/10.48550/arxiv.1903.05293}.  Among the myriad applications of these tests, placing bounds or constraints on well-motivated theoretical modifications to GR is the topic of this work~\cite{Yunes_2013,PhysRevD.86.044037}.

The action underlying the field equations of general relativity, the Einstein-Hilbert (EH) action, is a gem of predictive success, having survived a century of rigorous testing~\cite{Will_Yunes_2020,PhysRevD.55.4848}. However, one might expect that the EH action is not the whole story. For one, multiple candidates for theories of quantum gravity conjecture modifications to the EH action; the latter is linear in the curvature through the Ricci scalar, while (effective) quantum gravity models introduce curvature terms that are higher than linear~\cite{alexanderyunes2009,sopuertayunes2009,owen2021petrov}. Second, that the EH action produces a theory that is so successful across a wide range of energy scales suggests that any modification to GR may only appear at high curvatures; therefore, we might expect the EH action to merely be the leading-order term of an effective theory, expanded in powers of curvature~\cite{yunesstein,2009,dcs_detect}.  

In this work, we focus on two such modifications to GR that introduce quadratic curvature terms, dynamical-Chern-Simons (dCS)~\cite{alexanderyunes2009} and scalar-Gauss-Bonnet (sGB) gravity~\cite{Yunes:2011we,Yunes_2013}. These theories introduce a dynamical scalar (sGB) and a pseudo-scalar (dCS) degree of freedom that couples non-minimally to the metric through the Gauss-Bonnet (sGB) and Pontryagin (dCS) topological invariant, respectively. Both theories are well motivated, either from quantum gravity extensions of GR such as heterotic string theory~\cite{alexanderyunes2009,Cano_2019} or from effective field theories of gravity~\cite{Weinberg:2008hq,alexanderyunes2009}. These theories naturally avoid binary pulsar constraints~\cite{Yagi:2013mbt,yagi_challenging_2016}, but they are now beginning to be bounded through observations of gravitational waves with advanced LIGO (aLIGO) in the sGB case~\cite{constraints1,Perkins:2021mhb,Wang:2021jfc,Lyu:2022gdr,Wang:2023wgv} or observations of neutron stars with aLIGO and the Neutron star Interior Composition ExploreR (NICER) in the dCS case~\cite{constraints2}. 

Finding rotating BH solutions to modified field equations, such as in dCS and sGB gravity, is in general extremely complicated. One way to do so is through non-perturbative, numerical methods~\cite{Kleihaus_1996,Kleihaus_2011,Delsate:2018ome,Sullivan:2019vyi,Sullivan:2020zpf}, but, when doing so, one must be careful to properly resolve steep gradients, which may arise near horizons and curvature singularities. Another way to find modified BH solutions is perturbatively, as a simultaneous series expansion in small rotation or spin and in small (modified gravity) coupling~\cite{alexanderyunes2009,dcs2003,yagi_slowly_2012,Maselli_2015,Maselli_2017}. This is difficult because the perturbed modified field equations can become quite complicated at high enough spin order. Previous studies have often been stymied by this difficulty, and have therefore, until recently, been truncated to relatively low expansion order~\cite{inspiral2012,Deich_2022,dcs2003,owen2021petrov,cardenas-avendano_exact_2018,yagi_slowly_2012}. Such a truncation limits any study to only focusing on BHs with relatively low spin values, i.e.,~to BHs in a spin regime where the perturbation is valid. 

Recent work has made great progress in understanding how the modifications to the field equations produce modified BH solutions, allowing for analytic modified BH metrics of arbitrary order in a small spin expansion~\cite{Cano_2019}.  However, the high-order solutions generated through this procedure can be extremely cumbersome due to the high number of terms required in the expansions.  We then have a conundrum for the working physicist who wishes to use these modified metrics: if one wishes to calculate a given observable in a modified theory to some accuracy, what order of expansion should one use? While it would technically be feasible to simply use the highest-order solution possible, the unwieldy size of these solutions renders this route computationally impractical. Instead, it would be much better if one could calculate the sought-after observable using the BH metric at the lowest order needed for a given accuracy, leveraging the fact that all observations have finite statistical certainty, and thus, saving considerable computation time.

But how do we determine what this order is for a given modified BH metric? This is the main topic of this paper. We focus on BHs in dCS gravity and sGB gravity as an example and on the following ten observables: 
(i) the mass quadrupole moment; 
(ii) the photon ring perimeter radius; 
(iii) the angular momentum on the photon ring; 
(iv) the orbital frequency on the photon ring;
(v) the Lyapunov exponent on the photon ring; 
(vi) the perimeter radius of the ergosphere;
(vii) the perimeter radius of the innermost stable circular orbit (ISCO);
(viii) the angular momentum on the ISCO;
(ix) the binding energy on the ISCO; and (x) the orbital frequency on the ISCO.  For each of these observables, we calculate their corrections at each spin order in the metric up to order 24.  

Calculating these observables becomes tricky at high enough order in spin.  If done entirely analytically, the number of terms makes the calculations take an extremely long time, even on high-performance computing clusters.  On the other hand, if done entirely numerically, the precision required quickly overwhelms what is available with double precision. To overcome these issues, we have developed a novel, semi-analytic method where we calculate only the observable's \emph{correction} analytically, and then store the previous order value numerically. This allows us to perform the calculations in a reasonable amount of time, while also minimizing any numerical instabilities.

\begin{figure*}
  \centering
  \subfigure{\includegraphics[scale=0.5]{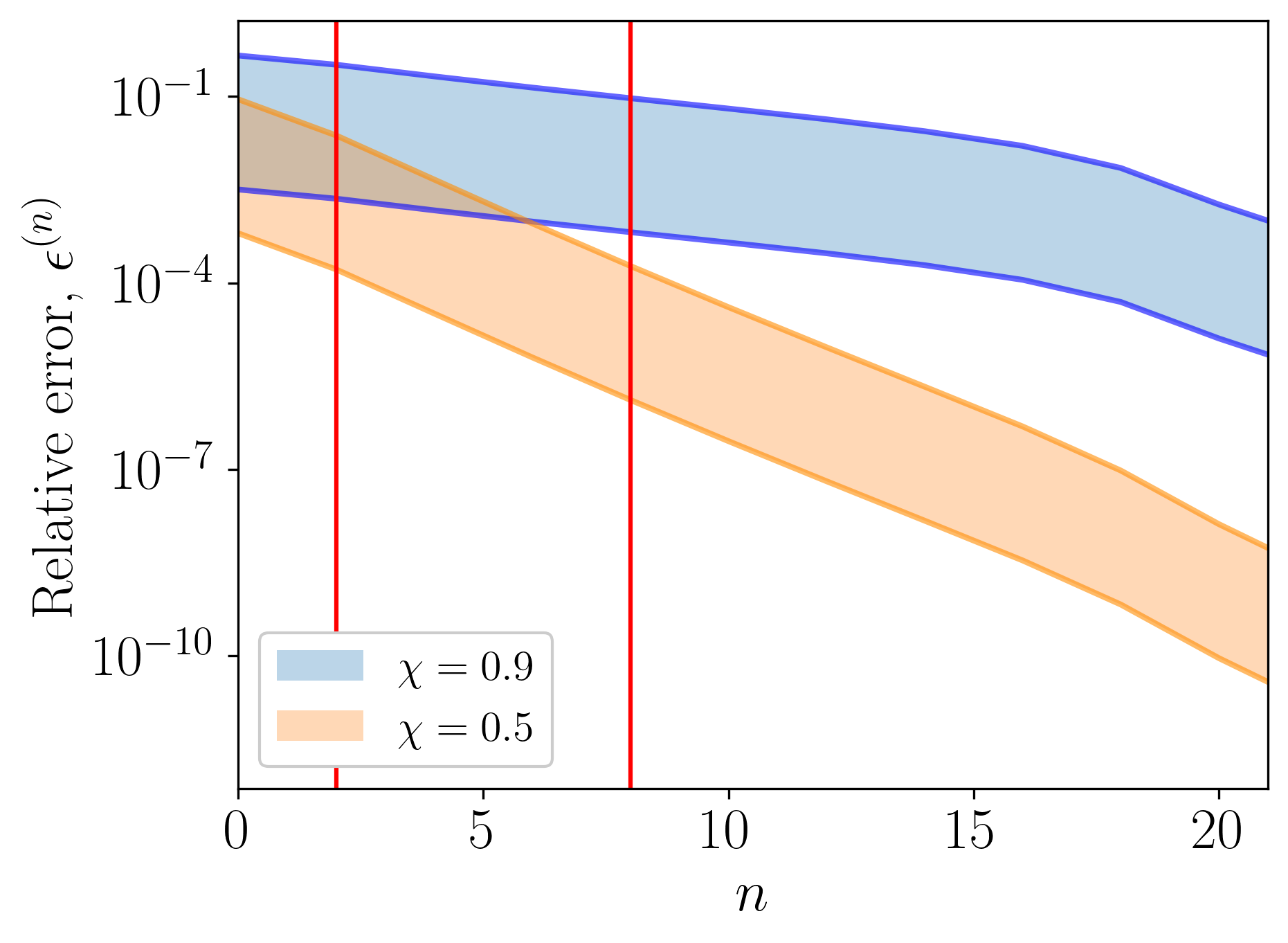}}%
  \quad\subfigure{\includegraphics[scale=0.5]{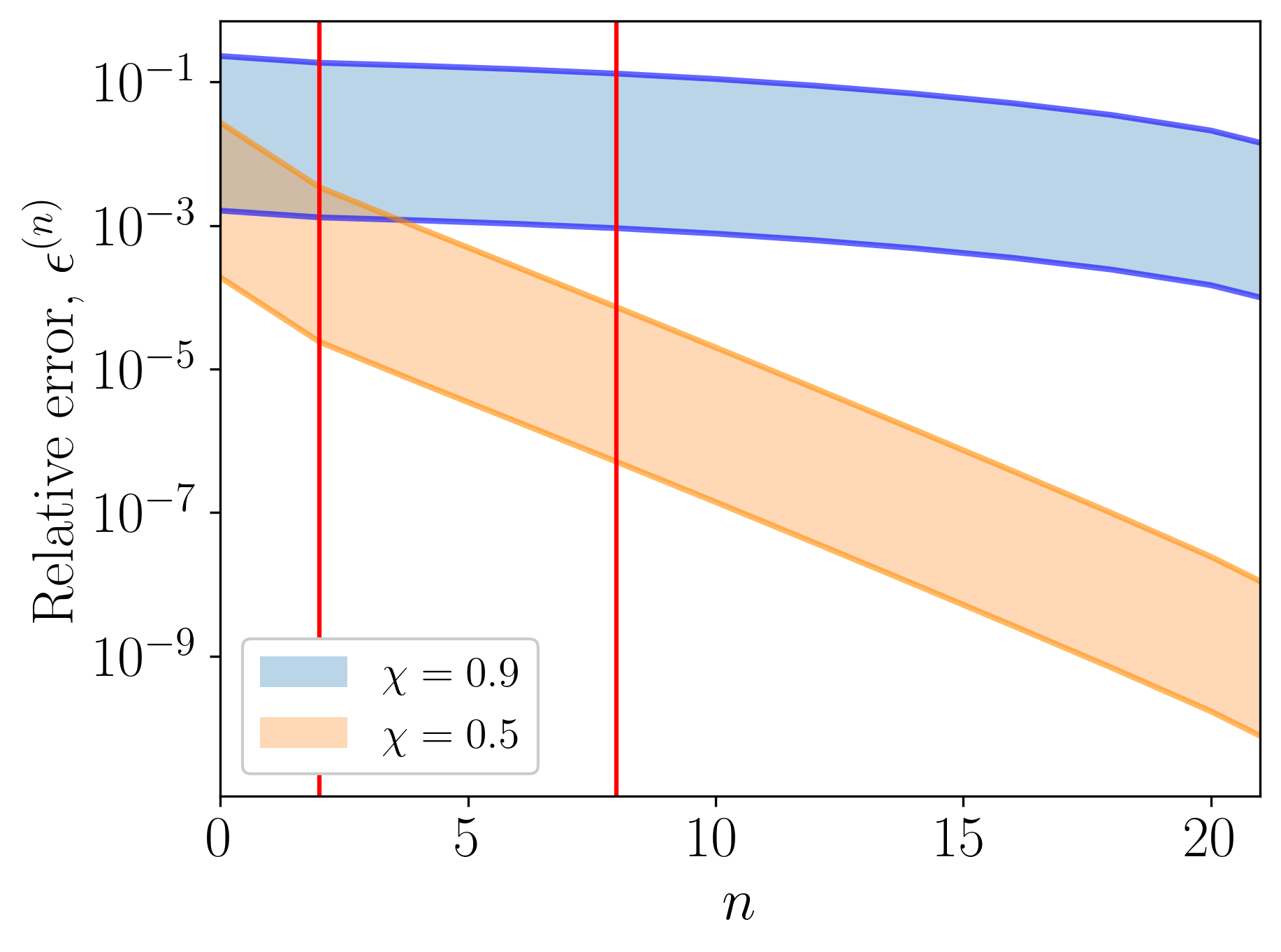}}
\caption{Spread of error in the calculation of all observables studied in this paper, as a function of the spin order kept in the approximate metric for dCS BHs (left) and sGB BHs (right) of two spin values and with the largest coupling allowed by the small-coupling approximation used to derive these metrics. Observe that an error of $\epsilon < 10^{-2}$ can be achieved for all observables with fewer than 8 terms in the expansion for BHs of moderate spin (red line on the right). This improves significantly for lower spin BHs and for smaller coupling parameters, allowing the same error to be achieved for spins less than $\chi=0.5$ with only 2 expansion orders (red line on the left).}

    \label{fig:errspread}
\end{figure*}

We find that for BH spin of less than 0.7, only the first 6 orders of spin expansion are required to calculate the observables to a relative difference of $10^{-2}$. Figure \ref{fig:errspread} shows how the error in the calculation of observables scales with the order kept in the spin expansion, for dCS BHs of various spins. Observe that only when the dimensionless spin is very high must a high order in the spin expansion be used. Throughout this paper, we will present the order required in the BH metric as a function of the BH spin value and the sought after accuracy.

Of course, ours is not the first work that has studied the required spin order in an approximate modified BH metric needed to place constraints on certain observables. Previous studies have performed similar analyzes, for example when looking at indicators of chaos in particle trajectories of these modified metrics~\cite{Deich_2022,cardenas-avendano_exact_2018} or when looking at BH shadow observations~\cite{SR_EM,bhshadow}. Our work extends these previous studies by not only considering far more observables, but also far higher orders in the small spin expansion, which have only recently been made possible by the (essentially) arbitrary-order metrics of~\cite{Cano_2019}. Our work, therefore, allows now for careful data analysis studies of these theories against observations that will be robust to the approximate nature of the modified BH metrics used.

The remainder of this paper presents the details that lead to the conclusions summarized above, and it is organized as follows. In Sec.~\ref{sec:solutions}, we discuss the two theories of quadratic gravity and their BH solutions.  In Sec.~\ref{sec:obsdef}, we give a description of the observables in question and describe how they are calculated. Section~\ref{sec:accuracy} covers how the error in the observables behaves as a function of spin and expansion order. Finally, Sec.~\ref{sec:conclusions} discusses the implications of the work presented.  Henceforth, we use geometric units in which $G=1=c$.

\section{Rotating Black Holes in sGB and dCS gravity}\label{sec:solutions}

Here, we present the basics of sGB and dCS gravity, and then briefly describe the BH solutions of each. 

\subsection{The quadratic gravity action}

We can modify the conventional EH action through an expansion in curvature terms, casting the EH term as merely the leading-order term in a broader effective field theory (EFT). In addition to this general EFT argument, quadratic theories also arise naturally from certain low-energy expansions of specific string theories~\cite{alexanderyunes2009,dcs2003}.

The action for these theories is defined as
\begin{equation}\label{eq:genaction}
    S = S_{\text{EH}} + S_{\text{mat}} + S_\vt + S_{RR},
\end{equation}
where $S_\text{EH}$ is the EH action, $S_\text{mat}$ is the matter action, $S_\vt$ is an action for a dynamical scalar or pseudo-scalar field, and $S_{RR}$ couples a quadratic curvature term to the field. The only distinction between the two quadratic theories we are concerned with is in this final term. The EH action reads
\begin{equation}
    S_\text{EH} = \kappa \int d^4x \, \sqrt{-g} \, R \,,
\end{equation}
where $\kappa = (16\pi)^{-1}$, $g$ is the metric tensor determinant and $R=g^{\alpha\beta}g^{\rho\sigma}R_{\rho\alpha\sigma\beta}$ the Ricci scalar, with the Riemann tensor $R_{\rho\alpha\sigma\beta}$. The action for the scalar or the pseudo-scalar field $S_\vt$ is 
\begin{equation}
\label{eq:action-sf}
    S_\vt = -\frac{1}{2}\int d^4x \sqrt{-g} \left[\nabla_\mu\vt\nabla^\mu\vt + 2V(\vt)\right],
\end{equation}
where $V(\vt)$ is the potential of the scalar field. In practice, we set $V(\vt)=0$ to ensure a massless theory~\cite{PhysRevD.93.029902}.

For the specific case of sGB and dCS gravity, we can write down a generic action that encompasses both theories after some parameter selection. Generically, we can say
\begin{equation}\label{eq:genquad}
\begin{aligned}
S_{RR} = \int d^4x\sqrt{\left|g \right|} \Big\{&\alpha_{\rm sGB} \vt_{\rm sGB} RR \\
&+ \alpha_{\rm dCS}\vt_{\rm dCS} R\tilde{R} \Big\},
\end{aligned}
\end{equation}

where
\begin{equation}
RR = R_{\mu\nu\rho\sigma}R^{\mu\nu\sigma}-4R_{\mu\nu}R^{\mu\nu}+R^2    
\end{equation}
 is the so-called Gauss-Bonnet density, and
 \begin{equation}
    R \tilde{R} \equiv {}^*R^\alpha{}_{\beta}{}{}^{\gamma\delta}R^{\beta}{} _{\alpha\gamma\delta}\,,
\end{equation}
is the Pontryagin density with ${}^*R^{\alpha}{}_{\beta}{}^{\gamma\delta} = \frac{1}{2}\epsilon^{\gamma\delta\rho\lambda}R^{\alpha}{}_{\beta\rho\lambda}$ the dual of the Riemann tensor. The parameters $\alpha_{\rm sGB}$ and $\alpha_{\rm dCS}$ determine the coupling parameter strength of the particular theory being described, and they have dimensions of length squared in geometric units.

From this generic non-minimal coupling action, we can now define sGB theory and dCS gravity. 
The action for sGB gravity is given by setting $\alpha_{\rm dCS} = 0$ in Eq.~\eqref{eq:genquad} and $\vartheta = \vartheta_{\rm sGB}$ in Eq.~\eqref{eq:action-sf}.  This finds motivation in a certain low-energy limit of string theory~\cite{Kanti:1995vq}. Gravitational wave observations have already constrained $\alpha_{\rm sGB}^{1/2} \leq 5.6\text{km}$ within a 90\% confidence interval~\cite{constraints1}. Unique among these two theories, sGB gravity induces modifications in the spacetime regardless of whether the spacetime is spherically symmetric (i.e., regardless of whether the BH is spinning or not).

On the other hand, dCS gravity is defined when $\alpha_{\rm sGB}=0$ in Eq.~\eqref{eq:genquad} and $\vartheta = \vartheta_{\rm dCS}$ in Eq.~\eqref{eq:action-sf}. In this case, $\vt_\text{dCS}$ behaves as a pseudo-scalar, on account of the fact that the Pontryagin density is odd under parity transformations. DCS gravity finds motivation from a few sources, including loop quantum gravity~\cite{dcs2003,yagi_challenging_2016}, the standard model gravitational anomaly~\cite{dcs2003,Yagi:2013mbt}, and investigations in string theory\footnote{A combination of both theories, sGB and dCS, with two scalar fields, also arises naturally in the effective action of heterotic string theory \cite{Cano:2021rey}.}~\cite{Campbell:1990fu,yagi_slowly_2012}.  Neutron star multi-messenger observations have been able to constrain $\alpha_{\rm dCS}^{1/2} \leq 8.5\text{km}$ 
 
within a 90\% confidence interval~\cite{constraints2}.  Unlike the sGB case, dCS gravity does not modify spherically symmetric spacetimes.  For this reason, it does not induce a change in a non-rotating BH. For a more thorough discussion, including the explicit field equations of both theories, see~\cite{Cano_2019}. 

In both cases, the modifications to the metric are proportional to the dimensionless coupling parameter
\begin{equation}
    \zeta_{\rm q} \equiv \frac{\alpha_{\rm q}^2}{\kappa M^4},
\end{equation}
which is the parameter we will use to present our results (where $\rm q$ stands for the theory being considered, either $\rm dCS$ or $\rm sGB$). In order to satisfy the requirement of the theories being effective, we assume the parameters $\zeta_{\rm q}$ are small, in a sense that we will make more precise below \cite{2021}.

\subsection{The corrected Kerr metric and its formal regime of validity}
Vacuum solutions in sGB and dCS gravity for axisymmetric and stationary spacetimes (in other words, the corrected forms of the Kerr metric in each theory) are found by starting with an ansatz for a corrected line element in Boyer-Lindquist-like coordinates $(t, r, \theta, \phi)$~\cite{Cano_2019}: 
\begin{widetext}
\begin{equation}
\begin{aligned}\label{eq:ansatz}
    ds^2 =& -\left(1-\frac{2M r}{\Sigma}-\zeta_{\rm q} H_1\right)dt^2
    -(1+\zeta_{\rm q} H_2)\frac{4M^2\chi r}{\Sigma}dtd\phi
    +\left(1+\zeta_{\rm q} H_3\right)\Sigma\left(\frac{dr^2}{\Delta}+\frac{dx^2}{1-x^2}\right)\\
    &+(1+\zeta_{\rm q} H_4)\left(r^2 + M^2\chi^2+\frac{2M^3\chi^2r\left(1-x^2\right)}{\Sigma}\right)(1-x^2)d\phi^2,
\end{aligned}
\end{equation}
\end{widetext}
with $\Sigma = r^2 + M^2\chi^2 x^2$ and $\Delta = r^2 +2Mr+ M^2\chi^2$. Here, $M$ is the mass of the BH, $\chi = a/M$ is the nondimensionalized spin, and $x=\cos\theta$.  The corrections $H_i$ are functions only of $r$ and $x$, and it is assumed that $|\zeta_{\rm q}H_i|\ll 1$ everywhere in the BH exterior. The quantity $H_i$  can be expressed as a power series in the spin:
\begin{equation}
\label{eq:H-exp}
    H_i = \sum_{n=0}^{\infty} H_i^{(n)}\chi^n,
\end{equation}
where the $H_i^{(n)}$ can always, for the theories under consideration, be written as a polynomial in both $1/r$ and $x$:
\begin{equation}
    H_i^{(n)} = \sum_{p=0}^{p_\text{max}} \sum_{k=0}^{k_\text{max}}H_i^{(n,p,k)}x^p r^{-k}.
\end{equation}
Here, $H_i^{(n,p,k)}$ are constant coefficients containing powers of $M$ and $k_\text{max}$ depends on $n$ and $p$. For a complete treatment, see~\cite{Cano_2019}.

\begin{figure*}
  \centering
  \subfigure{\includegraphics[scale=0.5]{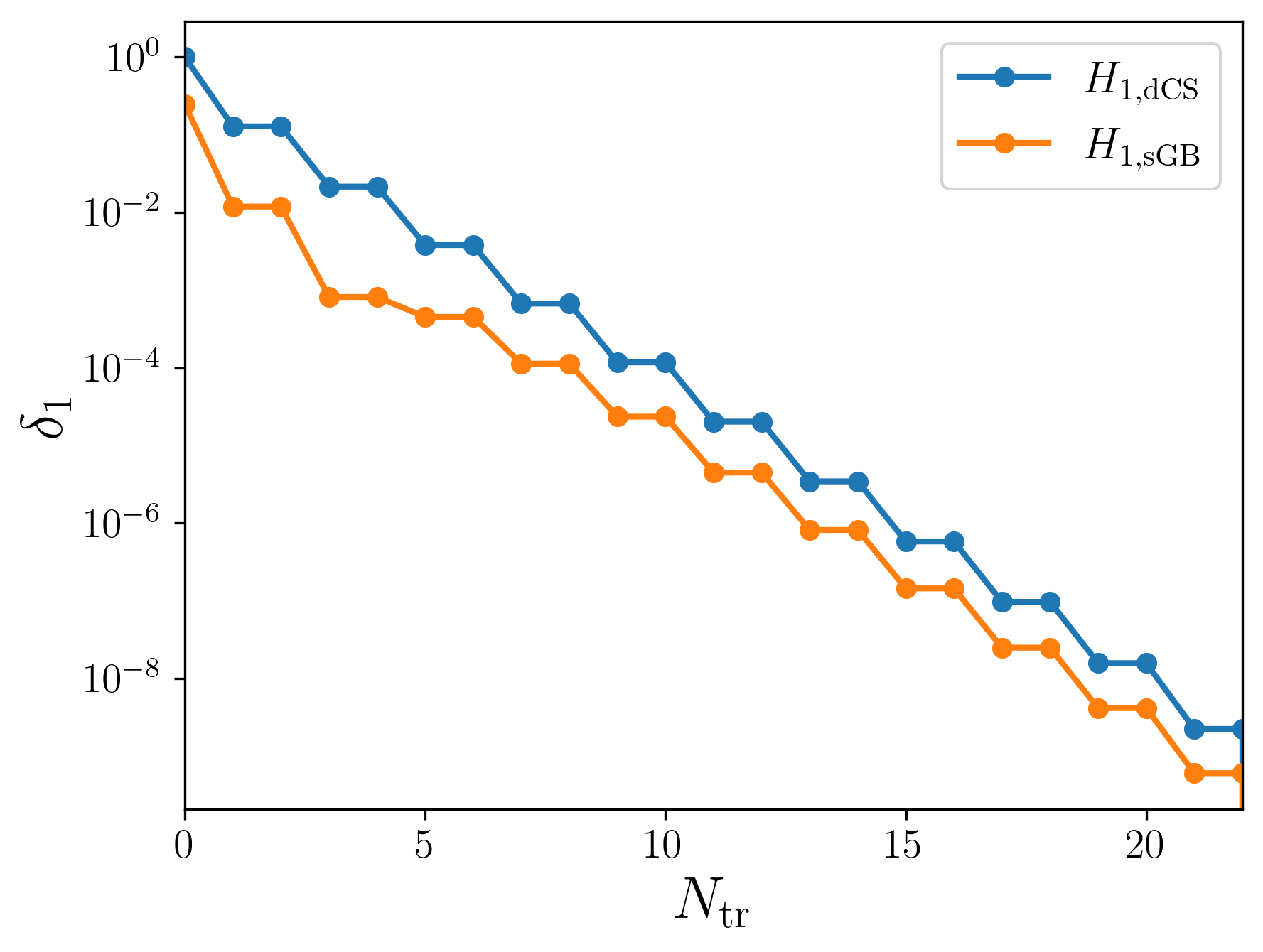}}%
  \quad\subfigure{\includegraphics[scale=0.5]{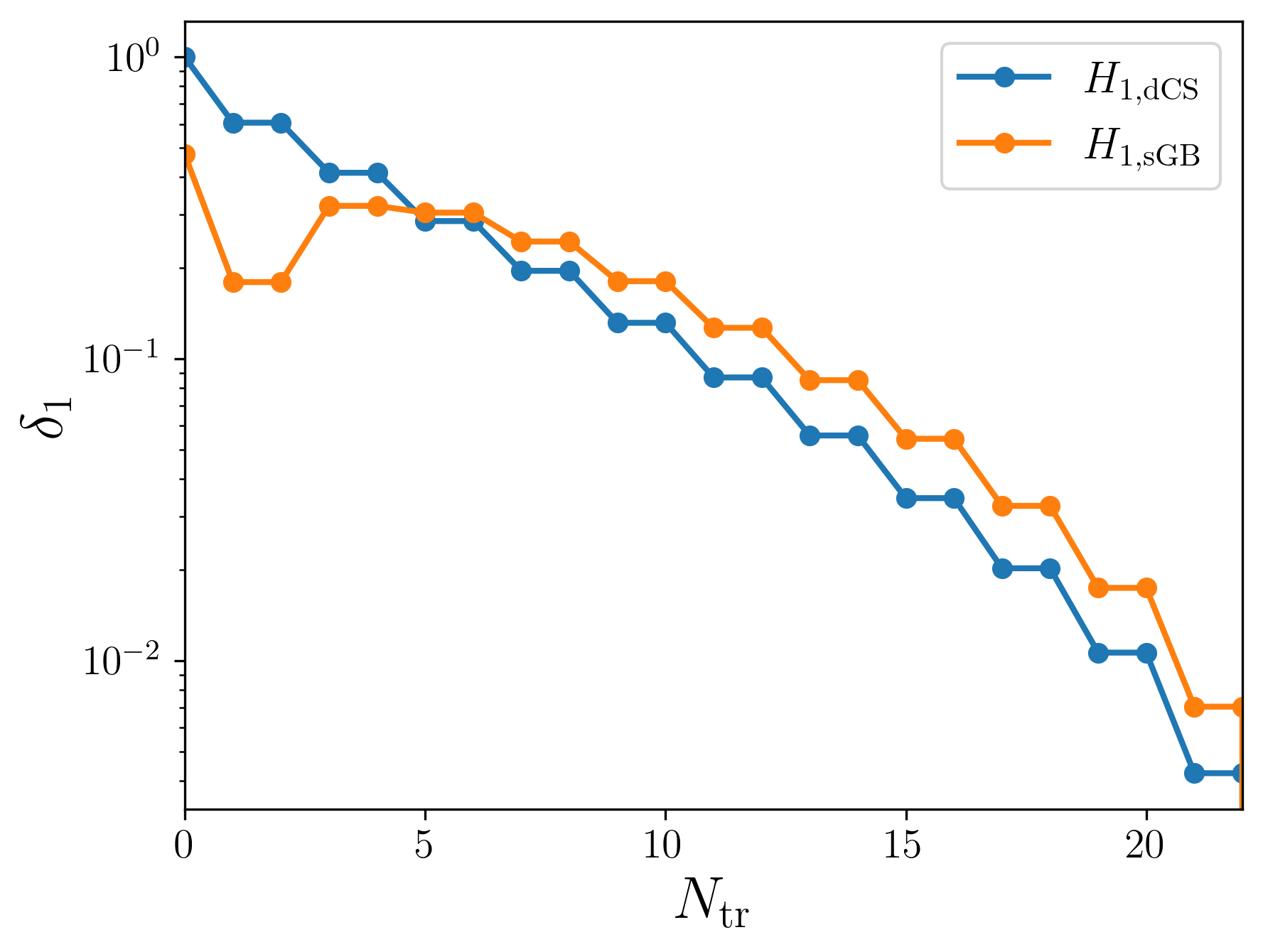}}
\caption{The $H_1$ correction functions for dCS (blue) and sGB (orange), evaluated at $r_\text{ISCO}$, with $\chi=0.4$ (left), and $\chi=0.8$ (right). These plots show the non-monoticity of the correction size for $H_{1,\text{sGB}}$ at low expansion orders and high spins.  Although the relative error of the $H_{i,\text{sGB}}$ do decrease at high expansion orders, this undesirable behavior should be noted for $N_\text{tr} < 6$. 
The other $H_i$ functions are superficially identical to the above.}

    \label{fig:h1example}
\end{figure*}

Using the corrections to the metric found in~\cite{Cano_2019}, we can now establish the highest allowed spin values at a given spin order from purely theoretical considerations. Let us define the relative error in the function $H_i$ via
\begin{equation}
    \delta_i \equiv 1 - \frac{\sum_{n=0}^{N_{\rm tr}} H_i^{(n)}\chi^n}{\sum_{n=0}^{N_{\rm hi}} H_i^{(n)}\chi^n}\,,
\end{equation}
where $N_{\rm hi}$ is the highest spin order considered in this paper. Figure~\ref{fig:h1example} shows this relative error for the $H_1$ function for various values of $N_{\rm tr}$. Observe that, while the dCS error monotonically decreases as the spin order is increased (even at high values of spin), this is not always the case for sGB corrections. At small values of $N_{\rm tr}$ and for large values of spin, the accuracy of $H_1$ does not improve monotonically, until $N_{\rm tr} > 6$. The slow error reduction of the sGB spin expansion limits us to using a maximum spin value of $\chi = 0.8$, while dCS allows calculations through to $\chi = 0.9$. 

We can now establish the maximum allowed value of the coupling parameters, $\zeta_{\rm sGB}^\text{max}$ and $\zeta_{\rm dCS}^\text{max}$, that are allowed by the perturbative solution. Recall that these approximate solutions are bivariate expansions in small spin (to higher order) and in the small coupling $\zeta_{\rm q}$ (to leading order), as explained around Eq.~\eqref{eq:H-exp}. We must therefore enforce that $|\zeta_{\rm q} H_i| \ll 1$, which we will use to find the maximum value of $\zeta_{\rm q}$ allowed. Evaluating the $H_i$ functions on the equator and at the event horizon with $\chi = 0.9$ in dCS gravity and $\chi = 0.8$ in sGB theory, and solving for the smallest value of the coupling parameter that makes one of the correction factors $|\zeta_{\rm q}H_i|$ larger than $0.5$, we find
\begin{equation}\label{eq:maxzeta}
    \zeta_{\rm sGB}^\text{max}~\approx~0.5\, ,\quad \zeta_{\rm dCS}^\text{max}~\approx~0.15\, .
\end{equation}
This clearly indicates that the $H_i$ functions (evaluated at the above values of $\chi$, on the horizon and the equator) are of order unity in both sGB and dCS gravity. Henceforth, we will limit our analysis to these maximum values of the dimensionless coupling, i.e. $\zeta_{\rm q} < \zeta_{\rm q}^{\rm max}$ and to $\chi \leq 0.8$ in sGB and $\chi \leq 0.9$ in dCS gravity. Just doing so, however, does not imply that observables calculated from these truncated approximate metrics will all be equally accurate, which is the topic of the rest of this paper.


\section{Definition of Observables}\label{sec:obsdef}
In this section, we describe every observable we will study in this paper. All observables are summarized in~\tab{tab:summary}.

\subsection{Multipole moments}
Multipole moments characterize the exterior field of a gravitating body. 
For stationary spacetimes in GR, they come in two classes: mass and angular multipoles, $M_l$ and $S_l$. For the Kerr BH, all of these are determined by the mass and angular momentum, and they satisfy
\begin{equation}
    M_{l}+i S_{l}=M(i a)^{l}\, .
\end{equation}
This is nothing but a manifestation of the no-hair theorems of GR, \textit{e.g.}~\cite{PhysRev.164.1776,PhysRevLett.26.331}. Hence, the measurement of at least one multipole moment, besides the mass and angular momentum, provides a test of the Kerr hypothesis. Multipole moments, in fact, leave an imprint in the inspiral phase of a compact binary \cite{PhysRevD.52.5707,Poisson:1997ha,Fransen:2022jtw}, and thus, on the gravitational waves that such a binary emits, allowing us to look for signatures of beyond-GR theories \cite{sopuertayunes2009,Yagi:2012vf,Wang:2021jfc,Perkins:2021mhb}. The mass quadrupole $M_2$ is the most relevant for observational purposes, so we focus on it here. 

The multipole moments can be identified by writing the metric in ACMC (asymptotically Cartesian and mass-centered) coordinates and reading-off certain terms in the $g_{tt}$ and $g_{t\phi}$ components \cite{RevModPhys.52.299}. This method was used in \cite{Cano:2022wwo} to compute the multipole moments in several higher-derivative theories, including dCS and sGB gravity. The value of $M_2$ for these theories reads
\begin{equation}
\begin{aligned}
    \frac{M_{2}^{\rm sGB}}{M^3}&=-\chi^2+\zeta_{\rm sGB}\left(-\frac{4463 \chi ^2}{2625}+\frac{33863 \chi ^4}{68600}+\ldots\right)\, ,\\
    \frac{M_{2}^{\rm dCS}}{M^3}&=-\chi^2+\zeta_{\rm dCS}\left(\frac{201 \chi ^2}{112}-\frac{1819 \chi ^4}{3528}+\ldots\right)\, ,
\end{aligned}
\end{equation}
and we have obtained the expansion of this quantity to order $\mathcal{O}(\chi^{28})$ for dCS gravity and to order $\mathcal{O}(\chi^{20})$ for sGB theory.  

\subsection{Geodesic Properties}

Here we describe the trajectory parameters we calculate, which are related to null and timelike geodesics. To facilitate our discussion of these trajectories, let us first review a few details about particle dynamics, which are common to all stationary and axisymmetric spacetimes. First, the stationarity and axisymmetry give rise to two Killing vectors, which in turn beget two conserved quantities of the motion.  Stationarity gives a conserved energy via the equation $E = -\mu^{-1} \xi^\nu_{(t)}p_\nu$, and axisymmetry gives a conserved angular momentum via $L = \mu^{-1}\xi^\nu_{(\phi)}p_\nu$, where $p_\nu = \mu u_\nu$ is the particle's 4-momentum (with $u_\nu$ its 4-velocity) and $\xi^\nu_{(X)}$ is the Killing vector associated with the $X$-direction~\cite{Misner1973}. A third conserved quantity is the rest mass of the particle itself, $\mu$, whose conservation is guaranteed by the conservation of the metric signature during the evolution of a geodesic.  In this work, $E$ (in the form of the binding energy, $E_b \equiv (\mu-E)/\mu$) and $L$ are crucial quantities we will derive from the approximate dCS and sGB metrics.

More precisely, the derivation of the $E_b$ and $L$ observables starts with the particular effective potential, $\veff$, that governs geodesic motion.  To derive the effective potential, we start with the normalization condition,

\begin{equation}
u^\alpha u_\alpha = -k,
\end{equation}
where $k=1$ for timelike geodesics parametrized by the particle's proper time and $k=0$ for null geodesics.
Then, making use of the conservation of energy and angular momentum, we can recast this equation in the particularly useful form~\cite{Deich_2022},
\begin{equation}
    \frac{1}{2}\left(u_r^2+u_\theta^2\right)+\veff=-\frac{1}{2}k,
\end{equation}
from which it can be shown that the effective potential $\veff$ is simply
\begin{equation}\label{eq:veff}
    \veff =\frac{1}{2}\left(\frac{g_{\phi\phi} E^2 + 2g_{t\phi} E L+L^2g_{tt}}{g_{tt}g_{\phi\phi} - g_{t\phi}^2} \right) ,
\end{equation}

Effective potentials can be analyzed in familiar ways from classical mechanics:  bound orbits correspond to the potential's extrema~(\fig{fig:veff_example}), and the potential's second derivative informs the orbit's stability to small perturbations. The potential is in general a function of $r$ and $\theta$, but here we only focus on equatorial trajectories, in which case the potential becomes a function of $r$ only.

\subsubsection{Photon rings}
The set of equatorial null orbits that are bound is called the \emph{photon ring}.  The radii and angular momenta corresponding to these orbits are found from the conditions
\begin{equation}\label{eq:prcond}
    \veff = 0 = \partial_r \veff,
\end{equation}
along with specifying that on the ring, $p_r = p_\theta = 0$~\cite{Misner1973,PhysRevD.5.814}.  All photon ring orbits have fixed $r$-coordinate values, which depend only on the BH spin. Photon ring locations are of particular interest to VLBI missions because they define the edges of the black hole shadow~\cite{Johnson_2020}.

\subsubsection{Lyapunov Exponents}
\begin{figure*}[htb]
  \centering
  \subfigure{\includegraphics[scale=0.5]{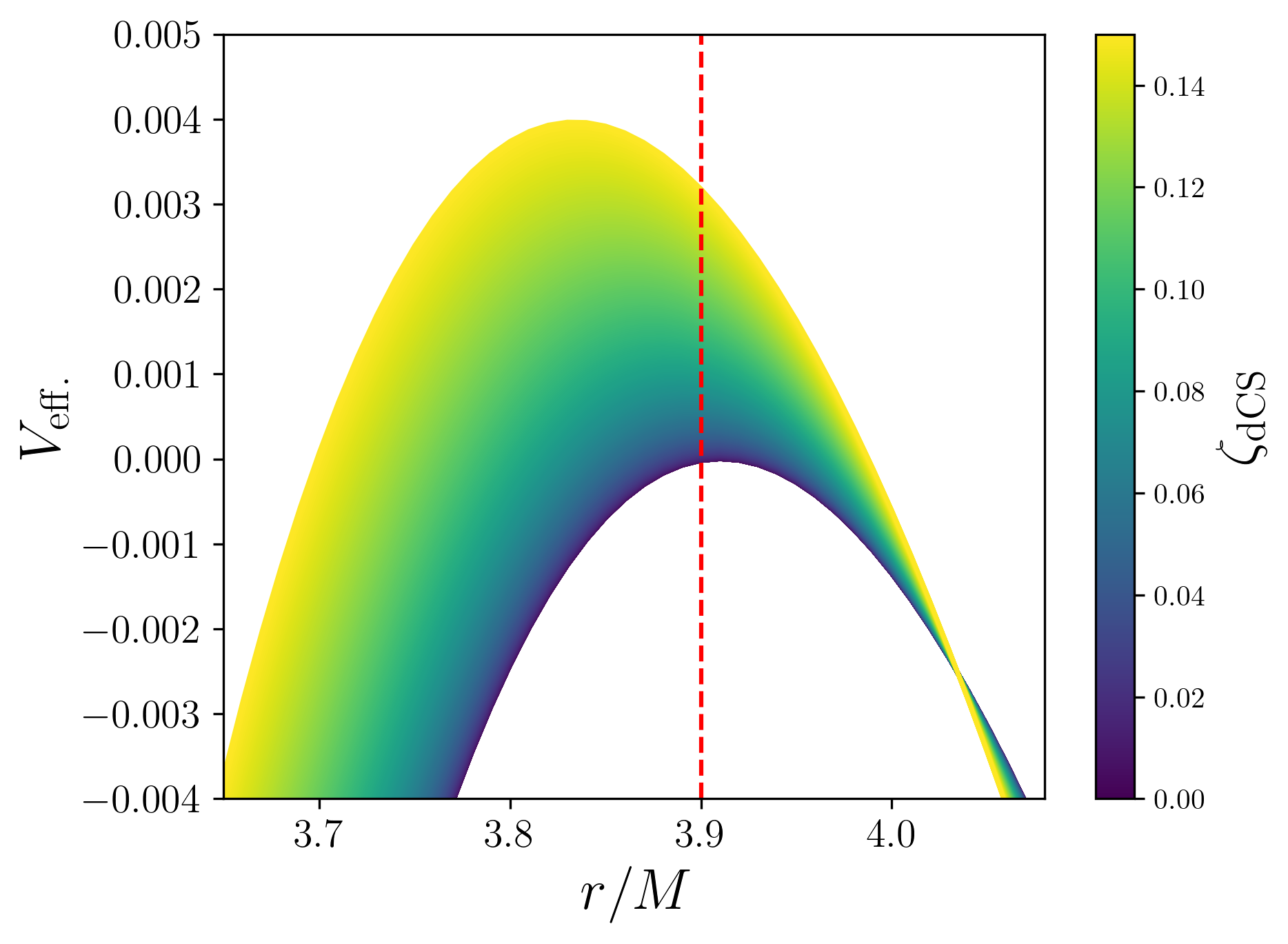}}%
  \quad\subfigure{\includegraphics[scale=0.5]{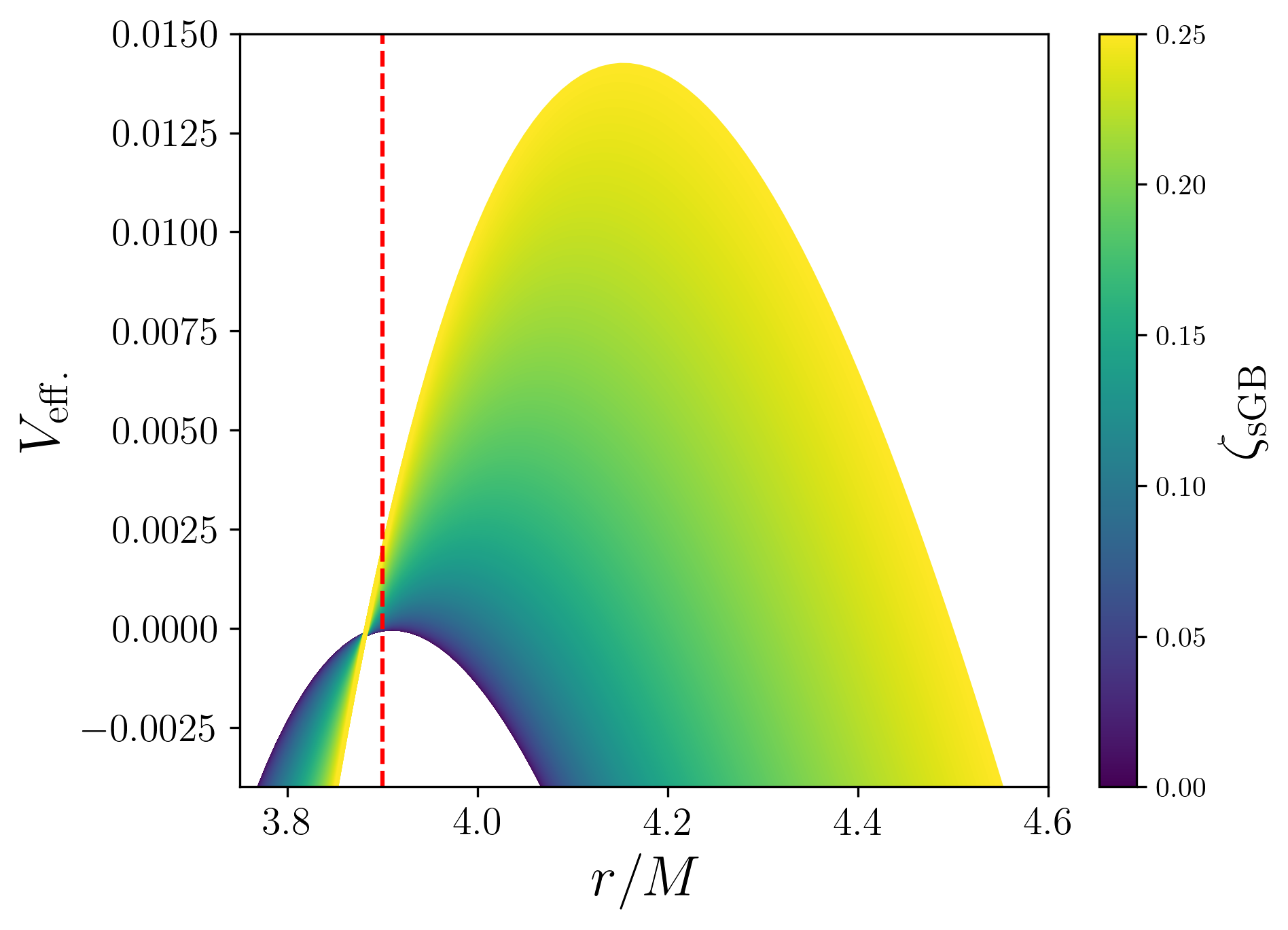}}
    \caption{A collection of null orbit effective potentials in dCS gravity (left) and sGB gravity (right) for varying values of $\zeta_\text{q}$, with $\chi=0.9$ and $L\approx-6.832$. The red lines at the peak of the $\zeta_\text{q} = 0$ curve correspond to the photon orbit in Kerr. Observe that this point moves as the coupling strength is increased.}
    \label{fig:veff_example}
\end{figure*}

 Owing to their position at the top of the ``hill'' of an effective potential~(\fig{fig:veff_example}), null bound orbits on the photon shell are inherently unstable to small perturbations, a fact that can be quantified through the orbit's \textit{Lyapunov exponents}. In general, for any dynamic Hamiltonian system, the stability of a phase space trajectory can be described by a Lyapunov exponent, $\lambda$.  While a full treatment of Lyapunov exponents of photon shell orbits does not permit analytic solutions, the symmetries of a photon ring orbit allow us to make significant simplifications.  An equatorial orbit has a two-dimensional phase space, in $r$ and $p_r$. This simplifies the form of the Lyapunov exponent significantly, which can be written as
\begin{equation}\label{eq:lyapdef}
    \lambda = \sqrt{\partial^2_r \veff}\,.
\end{equation}
While this form is sufficient for our present purposes, a full treatment of Lyapunov exponents in a generally relativistic context can be found in \cite{levin2000,Cornish_2001,Cardoso_2009}.  It should be noted that, in general, Lyapunov exponents in GR depend on the time parametrization of the given null geodesics.  The definition given in \eqref{eq:lyapdef} is implicitly in proper-time parametrization. Like photon ring locations, Lyapunov exponents are of interest for VLBI missions, as the Lyapunov exponent controls various aspects of the BH image, including magnification and ratio between fluxes of adjacent subrings \cite{Gralla_2020}.

\subsubsection{Innermost Stable Circular Orbit}
The ISCO is the timelike orbit around a BH with the smallest value of $r$ which is stable to small perturbations~\cite{Misner1973}.  The ISCO is an equatorial feature, being confined to the plane with $\theta = \pi/2$.  One can identify the ISCO by demanding the effective potential of a massive particle equal $-1/2$, along with its first two radial derivatives vanishing:
\begin{align}\label{eq:iscoconst}
    \veff = -1/2,\notag\\
    \partial_r \veff = 0 = \partial^2_r \veff.
\end{align}
ISCOs are of particular interest to observational BH astronomy, as they represent the inner boundary where light-emitting matter may be stably found.  For this reason, the ISCO is often considered the ``inner edge'' of astrophysical accretion disks~\cite{jefremov_innermost_2015,https://doi.org/10.48550/arxiv.2210.10500}.  For the ISCO, we will study the accuracy of the calculation of $r_\text{ISCO}$, $L_\text{ISCO}$, and $E_\text{ISCO}$, as well as $\omega_{\rm ISCO}$, the orbital frequency of massive particles at the ISCO. The latter is constructed from the Hamiltonian equations of motion, because $\omega_{\rm ISCO} = \phi'/t'$~\cite{amaro2018}, where primes indicate derivatives with respect to proper time. 

\subsubsection{Ergosphere}
The ergosphere is the region outside a BH in which it is impossible for a massive observer to remain stationary.  That is to say, the region where the four-velocity of a stationary observer becomes null, i.e. $g_{\mu\nu} u_{(t)}^\mu u_{(t)}^\nu > 0$, where $u^\alpha_{(t)}=dx^\alpha/dt = (1,0,0,0)$ is the tangent to the world line of a stationary observer. In practice, this implies solving $g_{tt} = 0$ for $r$~\cite{ayzenberg_slowly-rotating_2014,Misner1973}, which for Kerr reduces to the well-known result $r_\text{ergo} = M \pm \sqrt{M^2-a^2 \cos^2(\theta)}$. 
Rather than reporting the value of the $r$-coordinate for the photon ring, ISCO and ergosphere, we will focus on the ``perimeter radius'' $R = \sqrt{g_{\phi\phi}}|_r$, which represents the physical radius of each trajectory in the sense that $2\pi R$ is its arc length. 
\begin{table}[]
\caption{A summary of all observables studied in this paper for approximate metrics truncated at various orders in a small-spin expansion.}
\vspace{0.3cm}
\begin{tabular}{cc}
Symbol          & description 
 \\ \hline
$M_{2}$     & \begin{tabular}[c]{@{}c@{}}Mass quadrupole \\moment\end{tabular}
                    \\ \hline
$R_{\rm ph}$     & \begin{tabular}[c]{@{}c@{}}photon ring \\perimeter radius\end{tabular} \\ \hline
$L_{\rm ph}$     & \begin{tabular}[c]{@{}c@{}}angular momentum\\ on the photon ring \end{tabular}                       \\ \hline
$\omega_{\rm ph}$     & \begin{tabular}[c]{@{}c@{}}orbital frequency\\ on the photon ring \end{tabular}                       \\ \hline
$\lambda_{\rm ph}$       & \begin{tabular}[c]{@{}c@{}}Lyapunov exponent\\ on the photon ring\end{tabular}                      \\ \hline
$R_{\rm ergo}$       & \begin{tabular}[c]{@{}c@{}}perimeter radius\\of the ergosphere\end{tabular}  \\ \hline
$R_\text{ISCO}$ & \begin{tabular}[c]{@{}c@{}}perimeter radius\\of the ISCO\end{tabular}        \\ \hline
$L_\text{ISCO}$ & \begin{tabular}[c]{@{}c@{}}angular momentum\\ on the ISCO\end{tabular}                              \\ \hline
$E_b^{\rm ISCO}$           & \begin{tabular}[c]{@{}c@{}}binding energy\\ on the ISCO\end{tabular}                                \\ \hline
$\omega_{\rm ISCO}$     & \begin{tabular}[c]{@{}c@{}}orbital frequency\\ on the ISCO\end{tabular}                             \\ \hline
\end{tabular}
\label{tab:summary}
\end{table}

\section{Accuracy Required for Each Observable}\label{sec:accuracy}
In this section, we analyze the order in the spin expansion that is required to calculate several observables to a given accuracy. First, we outline the scheme by which we measure the errors as a function of expansion order, and then we summarize the results of calculating the errors for all quantities listed in \tab{tab:summary}.

\subsection{Error estimation scheme}
We will fix the modified gravity coupling constants $\zeta_{\rm q}$ to their maximum allowed values to ensure the approximate metrics remain valid outside the horizon [see Eq.~\eqref{eq:maxzeta}]. We will then study the relative difference between an observable computed with an ``exact'' metric and with a metric truncated at a given order in spin. By ``exact'' metric, we here mean the series-expanded metric truncated at a very high order in spin, such that the terms neglected introduce modifications below double precision. Since the series is convergent for $\chi < 1$ outside the horizon, taking the truncation above 20th order will suffice. In practice, we set the maximum truncation order, which defines an exact metric, to 24.

More precisely, any observable $A$ that depends only on the metric can be calculated to linear order in the deformation as
\begin{equation}
A = A_{\rm Kerr} +\zeta_{\rm q} \delta A  + {\cal{O}}(\zeta_{\rm q}^2)\,
\end{equation}
where $A_{\rm Kerr}$ is the observable computed with the Kerr metric (without expanding in small spins) and $\delta A$ is the deformation from Kerr introduced by the $\zeta_{\rm q}$-dependent terms in the metric. Since the modified gravity metric is known only as an expansion in small spins, the term $\delta A$ must also be cast as a series in spin, namely
\begin{equation}
    \delta A = \sum_{n}^N \delta A^{(n)} \chi^n + {\cal{O}}(\chi^{N+1})\,.
\end{equation}
In deriving these expressions, we have first expanded in small coupling $\zeta_{\rm q} \ll 1$, and then expanded only the deformation in small spins $\chi \ll 1$, without expanding the Kerr contribution in small spins, as this is known to all orders. 

With this in hand, we can define the relative fractional error between an observable computed with an ``exact'' metric $A_{\rm{ex}}$ and one computed with a truncated metric $A_{\rm{tr}}$ via

\begin{equation}
\label{eq:errdef}
\epsilon = 1 - \frac{A_{\rm tr}}{A_{\rm ex}}
= 
 1 - \frac{A_{\rm Kerr} +\zeta_{\rm q} \sum_{n}^{N_{\rm tr}} \delta A^{(n)} \chi^n}{A_{\rm Kerr} +\zeta_{\rm q} \sum_{n}^{N_{\rm hi}} \delta A^{(n)} \chi^n}
\,,
\end{equation}
where $N_{\rm hi} = 24$, while $N_{\rm tr}$ is a number we will vary. In what follows, we will study how the accuracy of the calculation of various observables changes as we increase $N_{\rm tr}$ toward $N_{\rm hi}$.

\begin{figure*}[htb]
  \centering
  \subfigure{\includegraphics[scale=0.45]{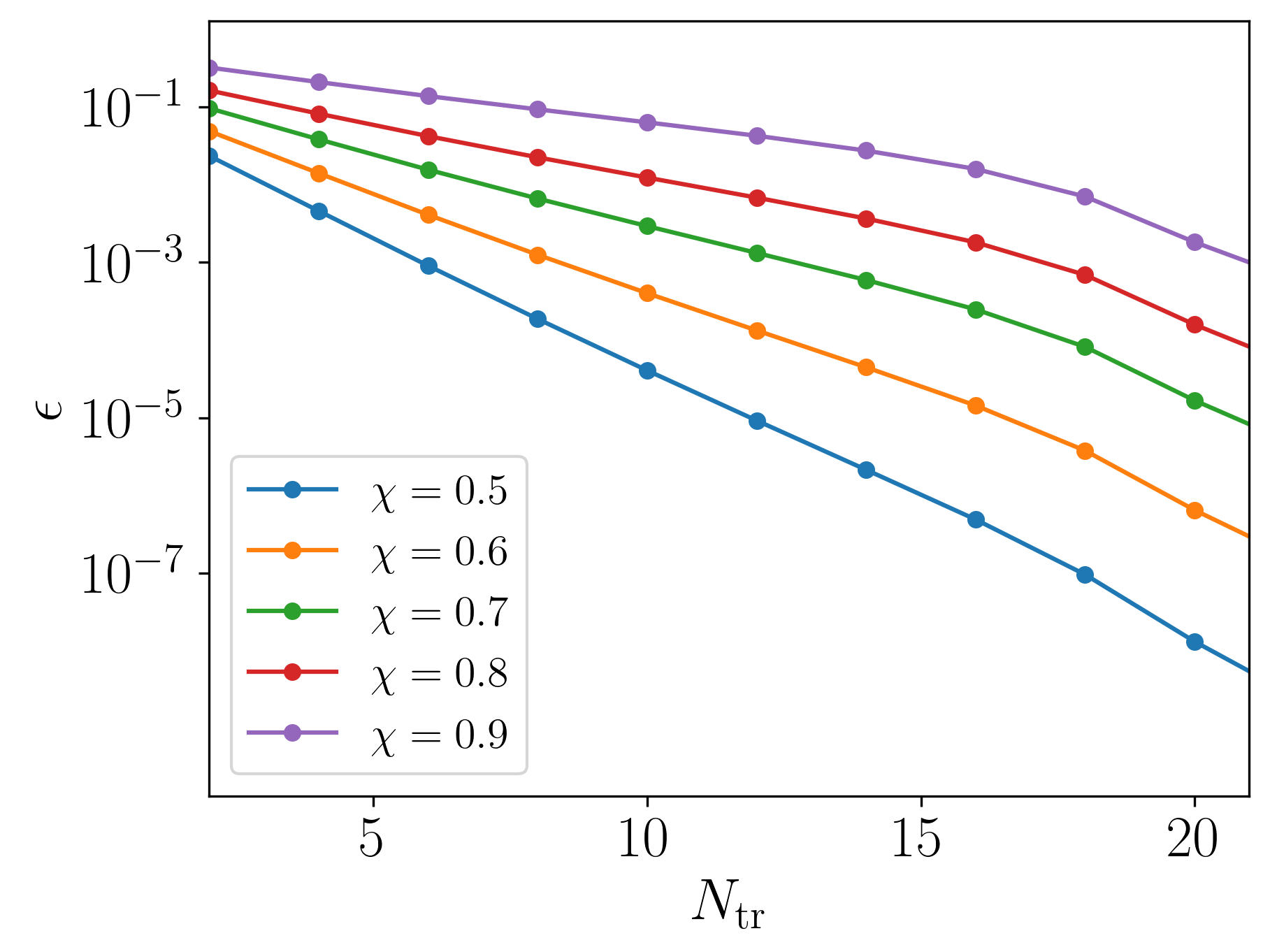}}%
  \quad\subfigure{\includegraphics[scale=0.45]{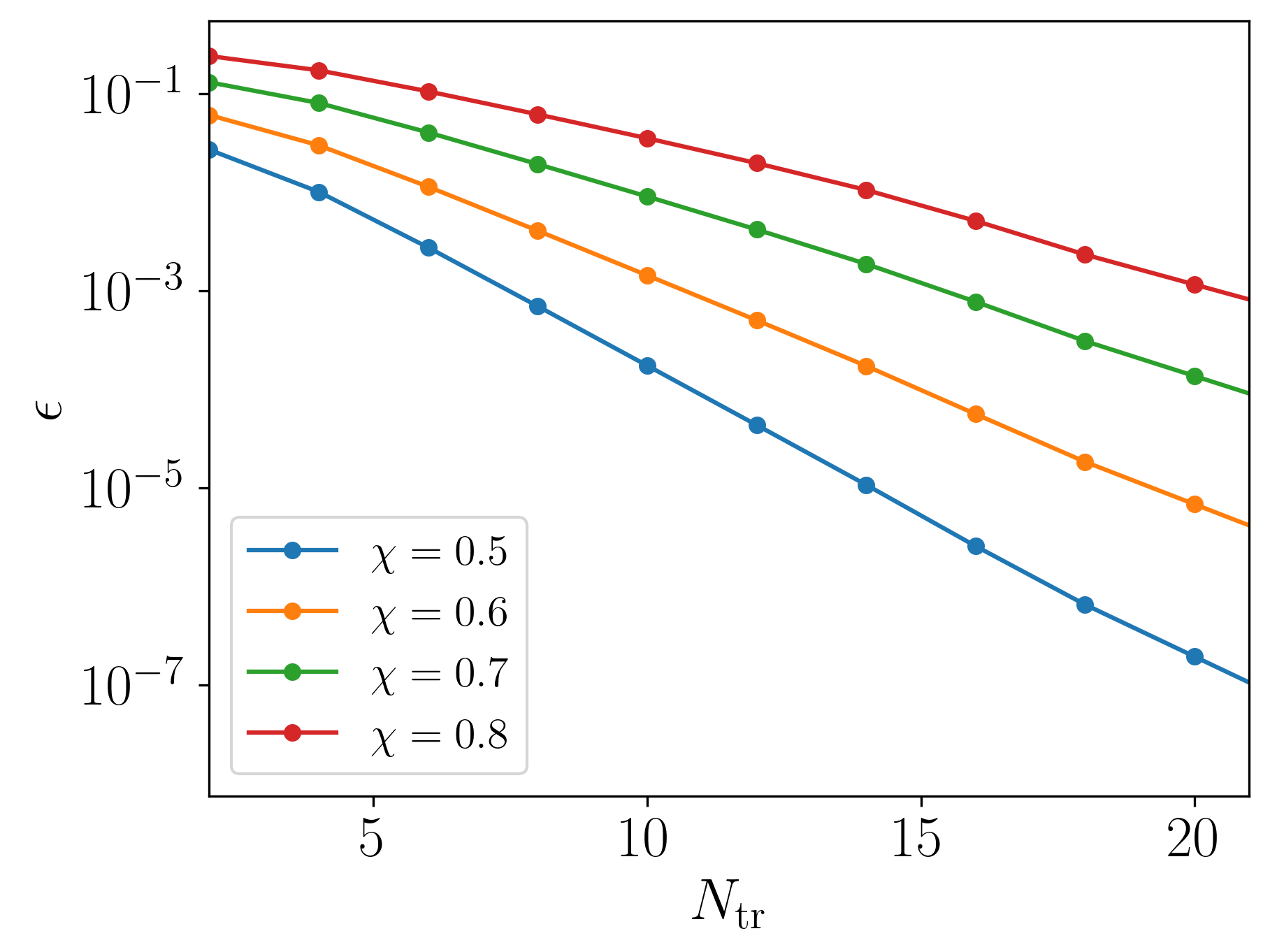}}
    \caption{Accuracy of the calculation of the Lyapunov exponent $\lambda_\text{ph}$, as a function of the truncation order of the metric in a slow-spin expansion. In this figure, we have saturated the coupling constants $\zeta_{\rm q}$ at the largest values allowed by the small-coupling approximation, with $\zeta_\text{dCS} = 0.15$ (left) and $\zeta_\text{sGB} = 0.5$ (right).
    }
    \label{fig:relerr}
\end{figure*}

\subsection{Implementation of the calculation of observables with a truncated slow-spin series}
We must be careful when calculating the observables described above in order to make their computation feasible. The reason for this is twofold.  If we try to perform the calculation entirely analytically, term-by-term, the computation becomes intractably slow for expansion orders greater than $N_\text{tr}\sim 10$, even on high-performance computing clusters. This is simply because the number of terms in the metric increases rapidly with the expansion order, making it extremely large beyond roughly ${\cal{O}}(\chi^{10})$. If, on the other hand, we try to perform the computation entirely numerically, we find numerically unstable behavior at large $N_\text{tr}$. This is because the $H_i$ terms, when expanded to order $N_\text{tr} > 10$, contain pieces that decay faster than $r^{-20}$ far from the black hole. These pieces are very small, even when evaluated close to the event horizon; for example, when evaluated at the ISCO, they are smaller than $10^{-16}$. The background metric and other pieces in the $H_i$ terms that are proportional to spin to a lower power, however, are of order unity. Therefore, when the small terms are added to the order-unity terms, one can quickly overwhelm double precision, yielding numerical calculations unstable.

To get around these issues, we here develop a semi-analytic approach, which allows us to limit the numerical noise, while also achieving useful speed.  The general idea is to always store the $(N_\text{tr}-1)^\text{th}$ term of the given observable to double precision numerically, and then use the perturbation to the metric to analytically calculate \emph{only the perturbation to the observable}.  For a concrete example, let us describe explicitly the process of calculating the first two corrections to the parameters of an ISCO orbit.

The location of the ISCO in Boyer-Lindquist coordinates $R_{\rm ISCO}$, the energy $E_{\rm ISCO}$, and the angular momentum $L_{\rm ISCO}$ are found simultaneously by solving the system of equations in~\eqref{eq:iscoconst}. Let us then define the following notation. First, we denote by $\mx_i \equiv (R_{\rm ISCO}, E_{\rm ISCO}, L_{\rm ISCO})$ a 3-vector containing the orbital parameters we are going to solve for. Then, we define $Y_i\equiv(\veff+1/2, \partial_r \veff, \partial_r^2 \veff)$ as a vector containing the symbolic effective potential (plus one-half), and its first two radial derivatives. With this notation, the location of the ISCO, its energy and angular momentum are simply found by solving the system of equations 
\begin{align}
\label{eq:master-ISCO}
Y_i(\mx_j) = 0\,,
\end{align}
for the $\mx_j$, which is simply a rewriting of Eq.~\eqref{eq:iscoconst} in more compact notation.

Let us now find these orbital parameters order by order. To start, we evaluate Eq.~\eqref{eq:master-ISCO}, 
\begin{equation}\label{eq:iscoconst2}
    Y^\text{Kerr}_i\Bigl(\mx^\text{Kerr}_i\Bigr) = 0,
\end{equation}
to find the usual $\mx_i^{\rm Kerr}$. To find the zeroth-order, the spherically-symmetric (i.e. $N_\text{tr} = 0$) correction, $\delta \mx^{(1)}_i$, we start by deriving the correction to the effective potential. At this order, the effective potential is
\begin{equation}
    \veff^\text{(1)} = \veff^\text{Kerr} + \delta \veff^{(1)},
\end{equation}
and $\delta \veff$ is the first-order perturbation:
\begin{equation}\label{eq:dveff}
    \delta \veff^{(1)} = \delta g^{(1)}_{tt} \frac{\partial \veff^\text{Kerr}}{\partial g_{tt}} + \delta g^{(1)}_{t\phi} \frac{\partial \veff^\text{Kerr}}{\partial g_{t\phi}} + \delta g^{(1)}_{\phi\phi} \frac{\partial \veff^\text{Kerr}}{\partial g^{(1)}_{\phi\phi}},
\end{equation}
where the $\delta g^{(1)}_{ij}$ terms are the $\mathcal{O}(\zeta_\text{q} \chi^0)$ perturbations to the metric given by Eq.~\eqref{eq:ansatz}. Equipped with the perturbation to the effective potential (and the corresponding vector $Y^{(10}_i = Y^\text{Kerr}_i + \delta Y^{(0)}_i$), we are now ready to calculate the correction to $\mx_i$, which we can find by linearizing Eq.~\eqref{eq:iscoconst2} about a small $\delta \mx_i$:
\begin{eqnarray}
    Y^{(1)}_i(\mx_i) &=& Y^\text{Kerr}_i\Bigl(\mx^\text{Kerr}_i+\delta\mx_i\Bigr)+\delta Y_i\Bigl(\mx^\text{Kerr}_i+\delta\mx_i\Bigr),
\notag \\
    &&\approx Y^\text{Kerr}_i\Bigl(\mx^\text{Kerr}_i\Bigr) + \delta \mx^j \left.\frac{\partial Y^\text{Kerr}_i}{\partial \mx^j}\right|_{\mx^\text{Kerr}_i} \notag \\
    &&+\delta Y_i\Bigl(\mx^\text{Kerr}_i\Bigr)+\delta \mx^j \left.\frac{\partial \delta Y_i}{\partial \mx^j}\right|_{\mx^\text{Kerr}_i},\\
    &&=\delta \mx^j\underbrace{\left.\frac{\partial Y_i^\text{Kerr}}{\partial \mx^j}\right|_{\mx^\text{Kerr}_i}}_{\text{(1)}} + \underbrace{\delta Y_i\Bigl(\mx^\text{Kerr}_i\Bigr)}_{\text{(2)}}.\label{eq:dxdef}
\end{eqnarray}

The first term of the second line vanishes by virtue of Eq.~\eqref{eq:iscoconst2}. Further, the last term of the second line is of second order in the perturbation, so we drop it for the calculation of $\delta \mx_i$. Then, in Eq.~\eqref{eq:dxdef}, term (2) is the perturbation to the effective potential vector, which is analytic and must be evaluated at the orbital parameters found in the previous iteration. The latter are known analytically in the Kerr case, but we will store them numerically, allowing us to also store term (2) numerically. Term (1) is the Jacobian of $Y^\text{Kerr}_i$ evaluated at the Kerr orbital parameters, which is calculated analytically, and then stored numerically. Finally, Eq.~\eqref{eq:dxdef} can be solved for the values of $\delta \mx_i$ that make this equation vanish. Before we move on to the next order, we update the Jacobian of term (1) by calculating $\partial \delta Y_i/\partial \mx^j$ analytically, and then evaluating it at the Kerr orbital parameters and storing the result numerically.

Let us now move to the $N_\text{tr} = 1$ correction. First, we find the effective potential as in the $N_\text{tr} = 0$ case, but now 
\begin{align}
    \veff^\text{(21} &= \veff^\text{Kerr} + \delta \veff^{(0)} + \delta \veff^{(1)}\,,
    \\
    &= \veff^\text{bg} + \delta \veff^{(1)}\,,
\end{align}
where we have absorbed the ${\cal{O}}(\zeta_{\rm q})$ correction into a modified background $\veff^\text{bg}$. The correction to this background effective potential, $\veff^{(1)}$, is then simply $\veff^{(0)}$ but with the replacements $\veff^\text{Kerr} \rightarrow \veff^\text{bg}$ and $\delta g^{(0)}_{ij}\rightarrow \delta g^{(1)}_{ij}$, where $g^{(1)}_{ij}$ is of ${\cal{O}}(\zeta_{\rm q} \chi)$. Making the same replacements in Eq.~\eqref{eq:dxdef} and using the same procedure we described above gives the corrections to $\delta \mx_i$ at second order. Clearly then, this method allows one to  bootstrap the correction to the ISCO observables (and to any other observable by following a similar procedure) to arbitrary order in perturbation theory, while minimizing the number of numerically troublesome terms.  At every step after $N_\text{tr}=0$, one must be careful to keep only the order in $\chi$ being considered, and expand all analytic expressions to $\mathcal{O}(\zeta_\text{q} \chi^{N_\text{tr}})$ whenever possible, to minimize numerical error.

\subsection{Accuracy of observables computed with a truncated slow-spin series}

Having laid out the scheme above, we are now ready to calculate the quantities described in Sec. \ref{sec:obsdef}. We here seek to understand how the relative error in the calculation of observables behaves as we vary the truncation order, for a given set of dimensionless spins and coupling constants. To get a grasp of this behavior, we present two types of figures that describe the error in the observables. The first type presents the relative error in Eq.~\eqref{eq:errdef} as a function of expansion order for several values of $\chi$, while setting $N_\text{hi} = 25$ and $\zeta_{\rm q}$ to the maximum values of Eq.~\eqref{eq:maxzeta}. 

Figure~\ref{fig:relerr} shows an example of this first type of figure for the calculation of the Lyapunov exponent, $\lambda_{\rm ph}$. Observe that as the truncation order is increased, the relative error decreases as expected. Observe also that we only plot even values of $N_{\rm tr}$. This is due to the fact that the $H_i$ functions only modify the metric at even orders in $\chi$, a fact already shown in \fig{fig:h1example}. Observe finally that as the spin increases, the number of terms needed to maintain a given accuracy also increases. For example, if one wishes to calculate $\lambda_{\rm ph}$ in dCS gravity to $0.1\%$ accuracy, then the Kerr deformations in the metric must be known to ${\cal{O}}(\chi^2)$ when the spin is $\chi < 0.4$, but the deformations need to be known to ${\cal{O}}(\chi^{20})$ if $\chi \sim 0.9$. SGB gravity presents similar behavior, but, in this case, even fewer terms are needed in the metric deformation to achieve a certain accuracy. At this point, we should also explain some unexpected behavior that appears in the relative error calculations for several observables in sGB gravity --- see the plots in Appendix \ref{sec:errplots}. The error is observed to increase briefly at low orders before continuing to reduce. This is simply due to the fact that the relevant calculated quantity changes sign, which can produce this behavior when the absolute value is taken, as is done for the calculation of $\epsilon$.

\begin{figure*}
  \centering
  \subfigure{\includegraphics[scale=0.5]{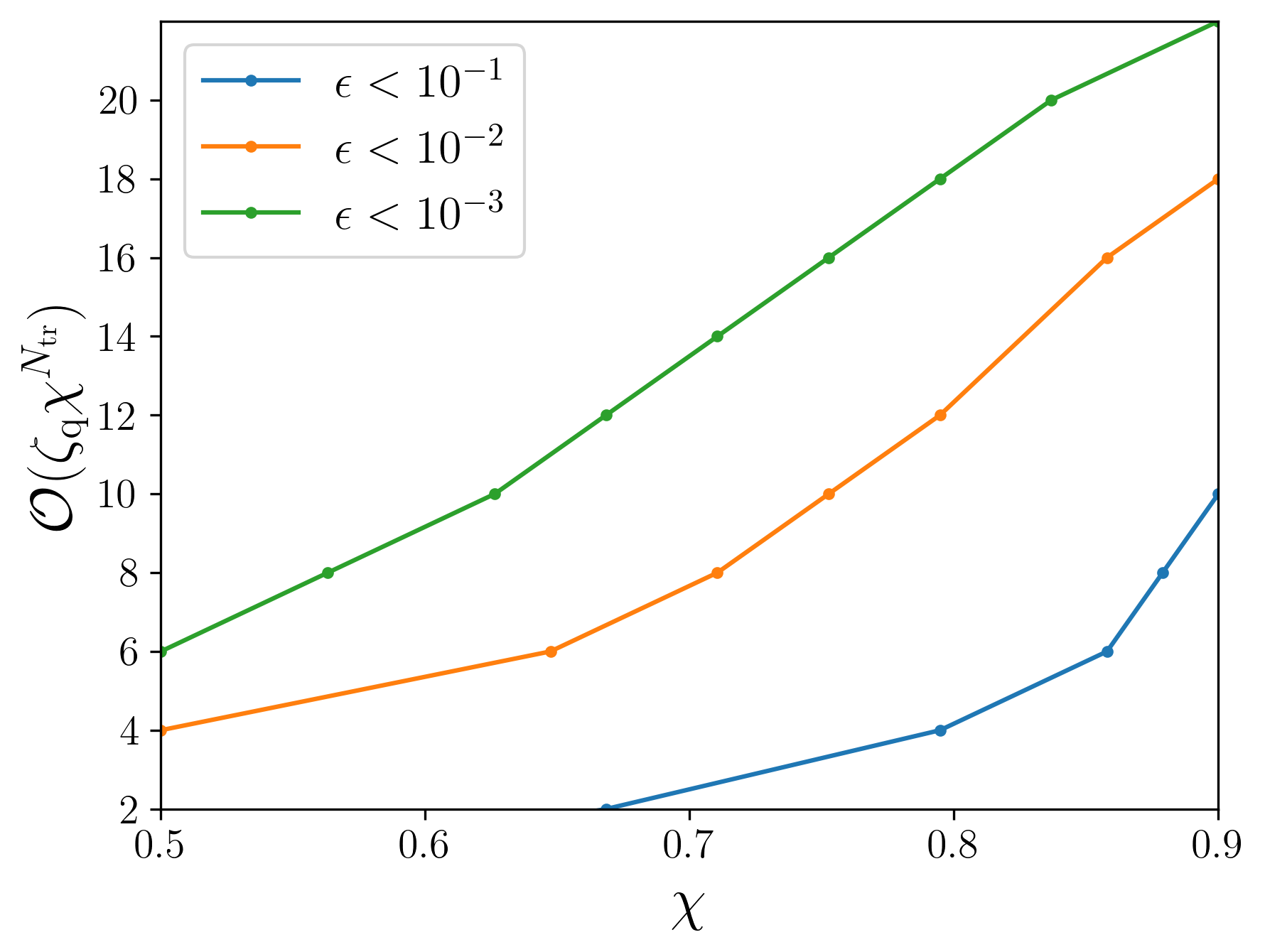}}%
  \quad\subfigure{\includegraphics[scale=0.5]{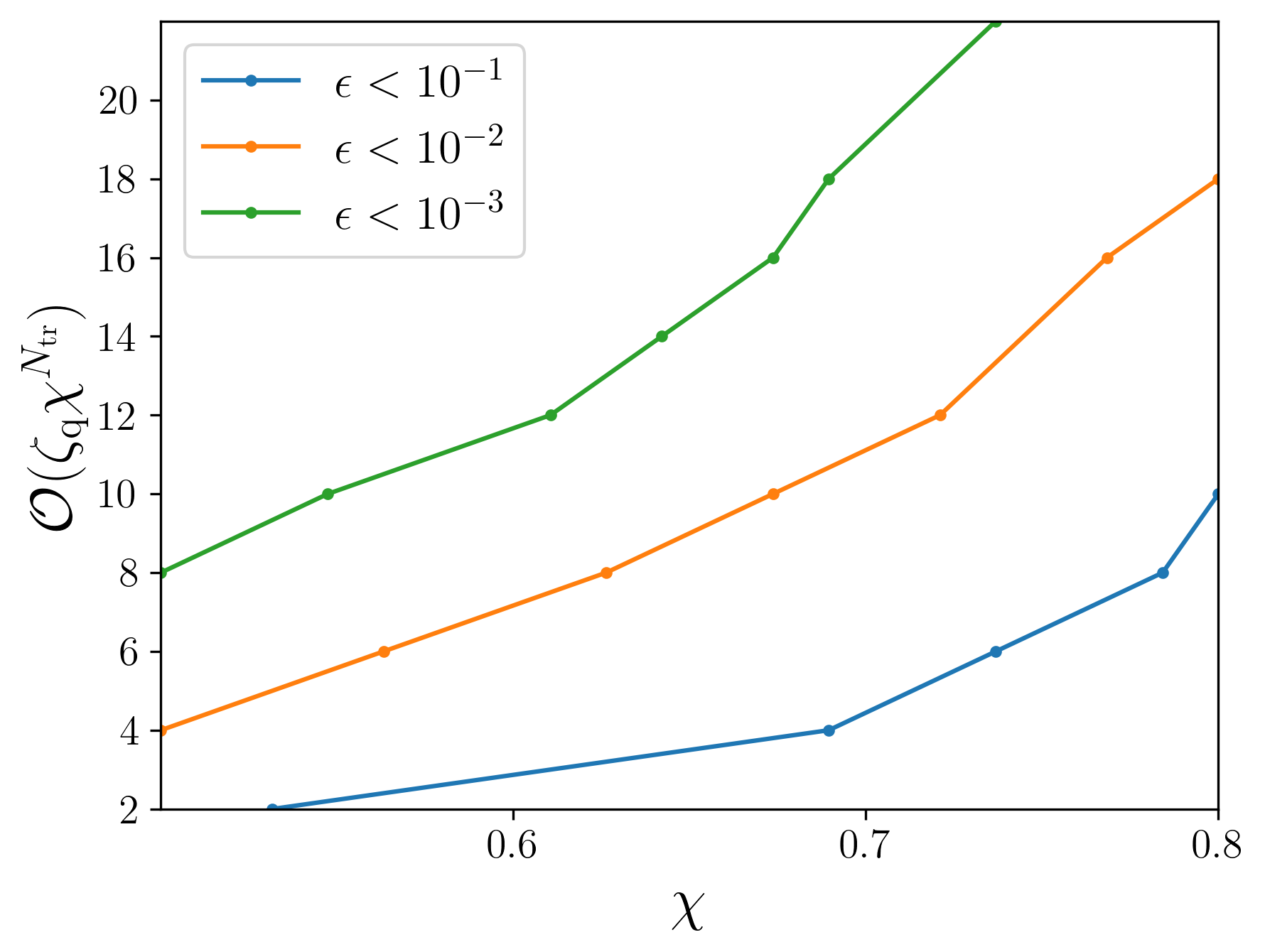}}
    \caption{The expansion order required to achieve a given relative error as a function of spin, for the Lyapunov exponent $\lambda_{\rm ph}$, with $\zeta_\text{dCS} = 0.15$ (left) and $\zeta_\text{sGB} = 0.5$ (right).}
    \label{fig:reqord}
\end{figure*}

The second type of plot is, in some sense, an inversion of the first type. For a given value of the coupling constants, we present the expansion order required to achieve an error less than a given threshold. Figure~\ref{fig:reqord} presents an example of such a figure, again focusing on the Lyapunov exponent, $\lambda_{\rm ph}$.  Observe that indeed, if the spin is low enough, then a few terms suffice to achieve good accuracy. For example, for an accuracy of $1\%$ in dCS gravity, the metric deformation must only be known to ${\cal{O}}(\chi^{5})$ if the spin $\chi < 0.6$. However, if one wishes to study more rapidly spinning BHs, such as one with $\chi = 0.9$, then the metric deformation must be known to ${\cal{O}}(\chi^{16})$. Observe that this is also the case for sGB gravity, although here fewer terms are needed. For example, for an accuracy of $1\%$ in sGB gravity, the metric deformation must only be known to ${\cal{O}}(\chi^{4})$ when the spin $\chi < 0.6$. 

The other observables we presented in Sec.~\ref{sec:obsdef} show very similar behavior. In order to avoid cluttering the paper, we have included these figures in Appendix~\ref{sec:errplots} and~\ref{sec:reqord}. To summarize all results, Fig.~\ref{fig:errspread} presents the spin order to which different metric deformations need to be kept to achieve a given accuracy, for BHs of various spins. Each band presented in that figure corresponds to the set of curves for all observables studied, at a fixed value of spin. For example, the blue-shaded region in Fig.~\ref{fig:errspread} corresponds to the region occupied by the relative accuracy curves for all observables of Table~\ref{tab:summary}, corresponding to the blue curves in the plots of Appendix~\ref{sec:errplots}. Observe that, once more, the number of terms required to obtain $1\%$ accuracy is rather small for spins $\chi < 0.5$, but for more rapidly spinning BHs, more terms need to be kept.

These plots make it clear that for $\chi \lesssim 0.6$, corrections of $\mathcal{O}(\zeta_{\rm q} \chi^8)$ provide excellent accuracy, with a relative difference $\epsilon$ being less than $10^{-4}$ in almost all cases of the coupling parameter~(\fig{fig:errspread}). For intermediate spins, with $0.6 < \chi < 0.8$, the story is more complicated, depending on the observable in question. For high spins of $\chi > 0.8$, the accuracy increases much more slowly.

\section{Conclusions}\label{sec:conclusions}

We have here calculated several observables associated with the spacetime outside BHs described by two different quadratic theories, sGB and dCS gravity. These BH metrics are now known to linear order in the coupling but to very high order in a small spin expansion, begging the question of what order must be kept in the latter to obtain sufficiently accurate calculations. The answer to this question is non-trivial because various observables that one may wish to calculate depend non-linearly on the various components of the metric. We have here carried out a careful and extensive exploration of the accuracy in the calculation of various observables and showed that, for large regions in spin and coupling parameter space, only relatively few orders of expansion in spin are required to achieve quite impressive accuracy. 

We expect these results to be of interest when computing other quantities with direct observational interest in detectors, such as quasinormal modes, gravitational waveforms and shadows. Since the computation of these more realistic observables can be challenging, it is crucial to reduce the computational cost by truncating the spin expansion of the metric at the minimum order required to obtain a given desired accuracy. The work presented here provides a suggestion for what this minimum order should be. Future work could study whether indeed the order requirements suggested here for a given set of observables also applies to other observables of interest.  

For instance, photon ring properties are sometimes related to the quasinormal mode spectrum of BHs through the geodesic analogy \cite{Cardoso_2009,Konoplya:2017wot,Konoplya:2022gjp}, even in some modified gravity theories~\cite{Glampedakis_2019}. Future experiments may be able to measure these modes with an accuracy of around $1\%$~\cite{Baibhav_2020,Bhagwat:2023jwv}. Thus, to be on the safe side, one should compute the QNMs with an accuracy higher than this. Using our results for the photon ring frequency as a reference --- see Figs.~\ref{fig:relerror1} and \ref{fig:requiredorder1} --- if one wishes to calculate QNMs with an accuracy of $1\%$ for a BH of $\chi=0.7$, one would need expansions of at least order $N_{\rm tr, sGB}\sim N_{\rm tr, dCS}\sim 6$. On the other hand, we need expansions of around twice this order in order to achieve an accuracy of $0.1\%$. One could therefore attempt to compute quasinormal modes with such a metric and determine whether indeed this order in small-spin expansion is sufficient. Of course, BH perturbation theory for small-spin metrics can itself be a very challenging calculation, with results only known to first~\cite{Pierini:2021jxd,Wagle:2021tam,Srivastava:2021imr}  or second~\cite{Pierini:2022eim} orders in spin in dCS and sGB gravity --- see also \cite{Cano:2021myl,Li:2022pcy,Hussain:2022ins,Cano:2023tmv}. Other, numerical methods for QNM calculations, such as wave-packet scattering, may be more suitable for calculations with higher-order-in-spin metrics.

Further, there is more exploration to be done of other high-curvature modifications to gravity, such as those with quartic~\cite{Endlich:2017tqa,Sennett_2020} and cubic \cite{Cano_2019} terms in the curvature, Einstein--\AE{}ther theory~\cite{Jacobson_2001}, or even other coupling functions in Gauss-Bonnet gravity that are not merely the shift-symmetric version considered here, such as in~\cite{Clifton_2012}. Finally, it would be useful to confirm the results found here through a parameter estimation study of a set of observables. One could imagine a study wherein GRMHD simulation data was compared to real-life interferometric data in order to perform a similar error analysis to what has been presented here. Our study lays the foundations for such follow-up work.

\vspace{0.4cm}
\begin{acknowledgments}  
The work of PAC is supported by a postdoctoral fellowship from the Research Foundation - Flanders (FWO grant 12ZH121N). NY and AD are supported through the Simons Foundation Award No.~896696 and the NSF Award PHY-2207650.
\end{acknowledgments}

\appendix
\section{Relative error plots}\label{sec:errplots}

Below we show the relative error plots for every observable listed in~\tab{tab:summary}.
\begin{figure*}[p]
    \centering
    \includegraphics[scale=0.425]{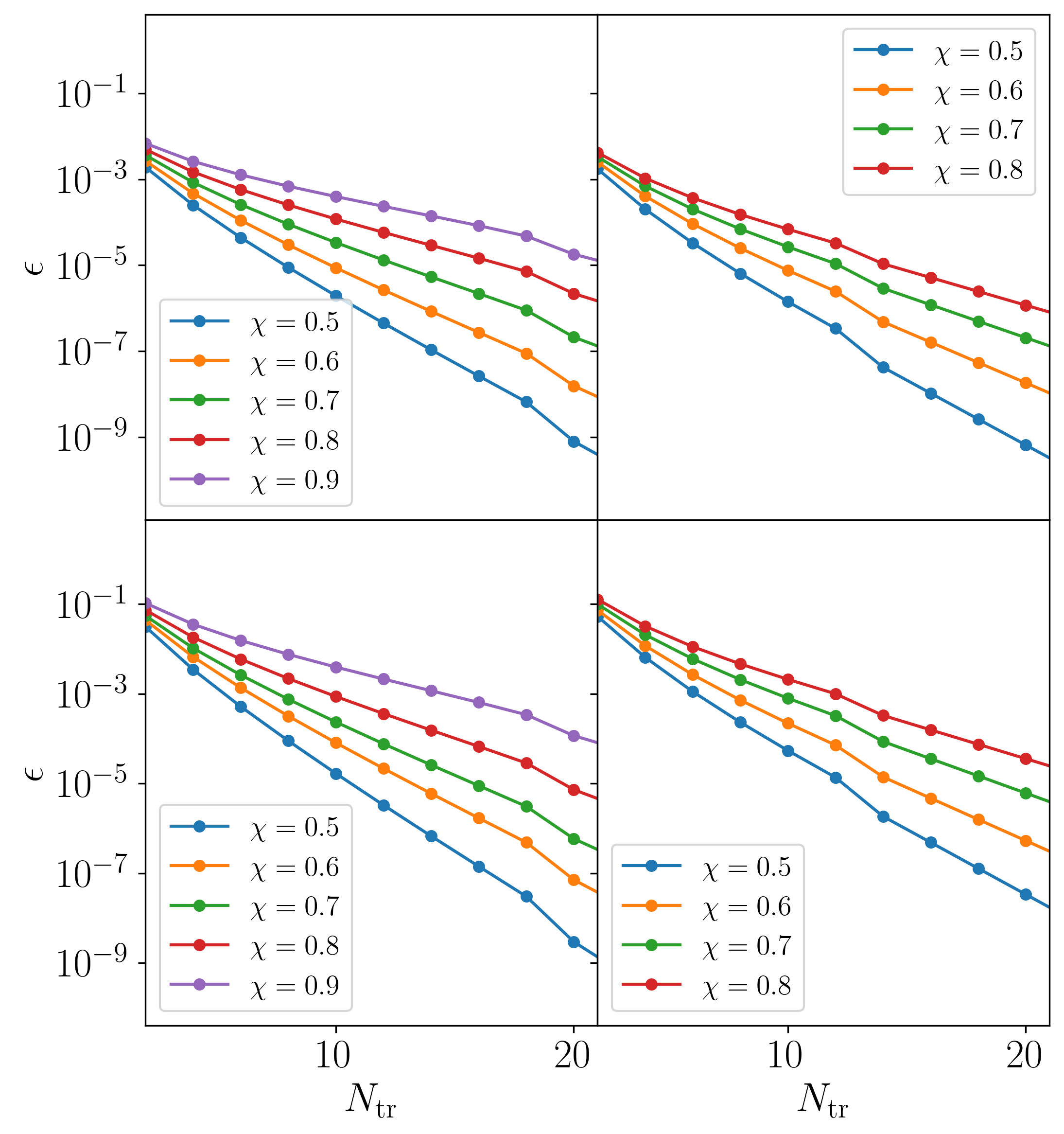}
    \includegraphics[scale=0.425]{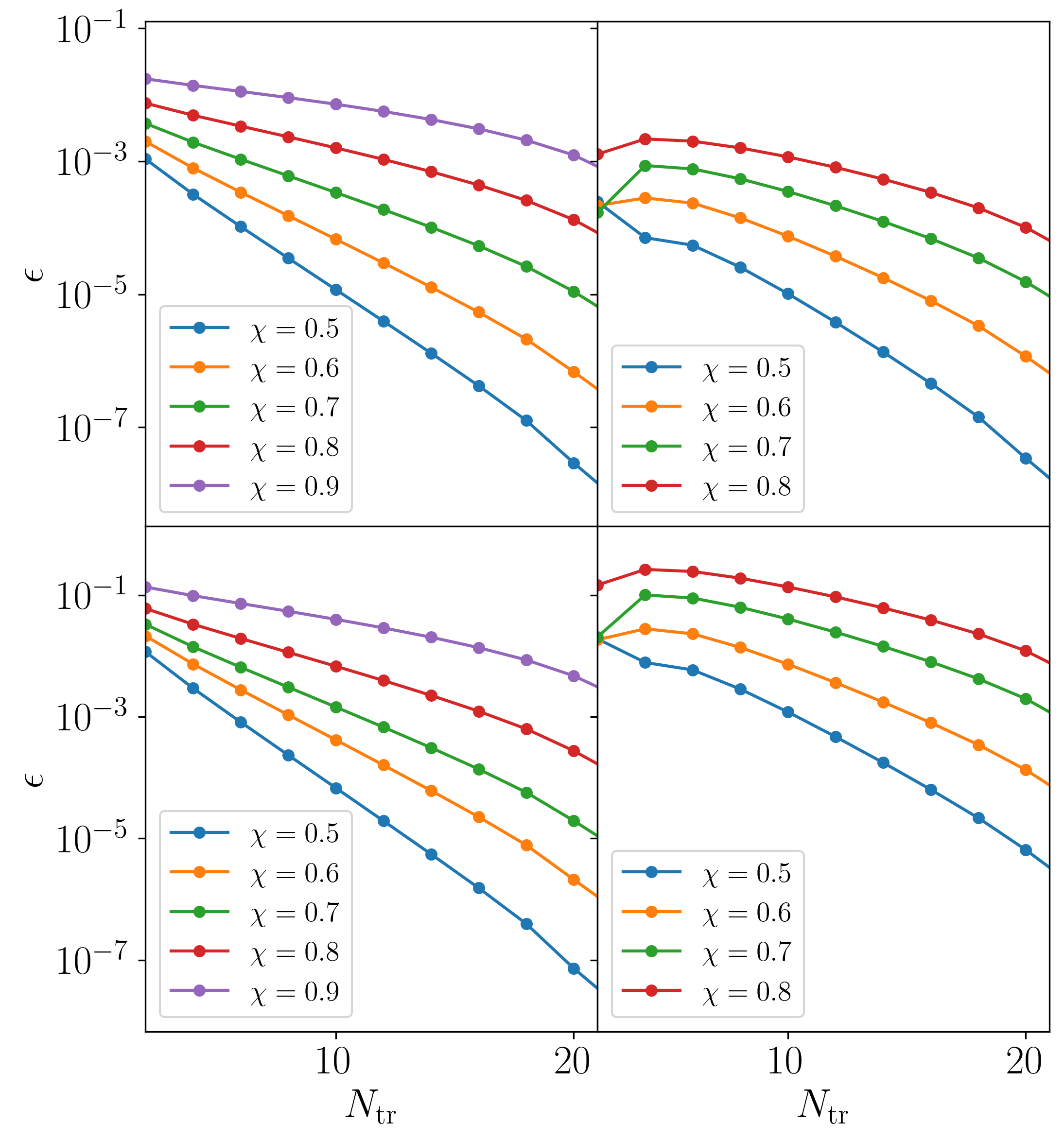} \\
    \includegraphics[scale=0.425]{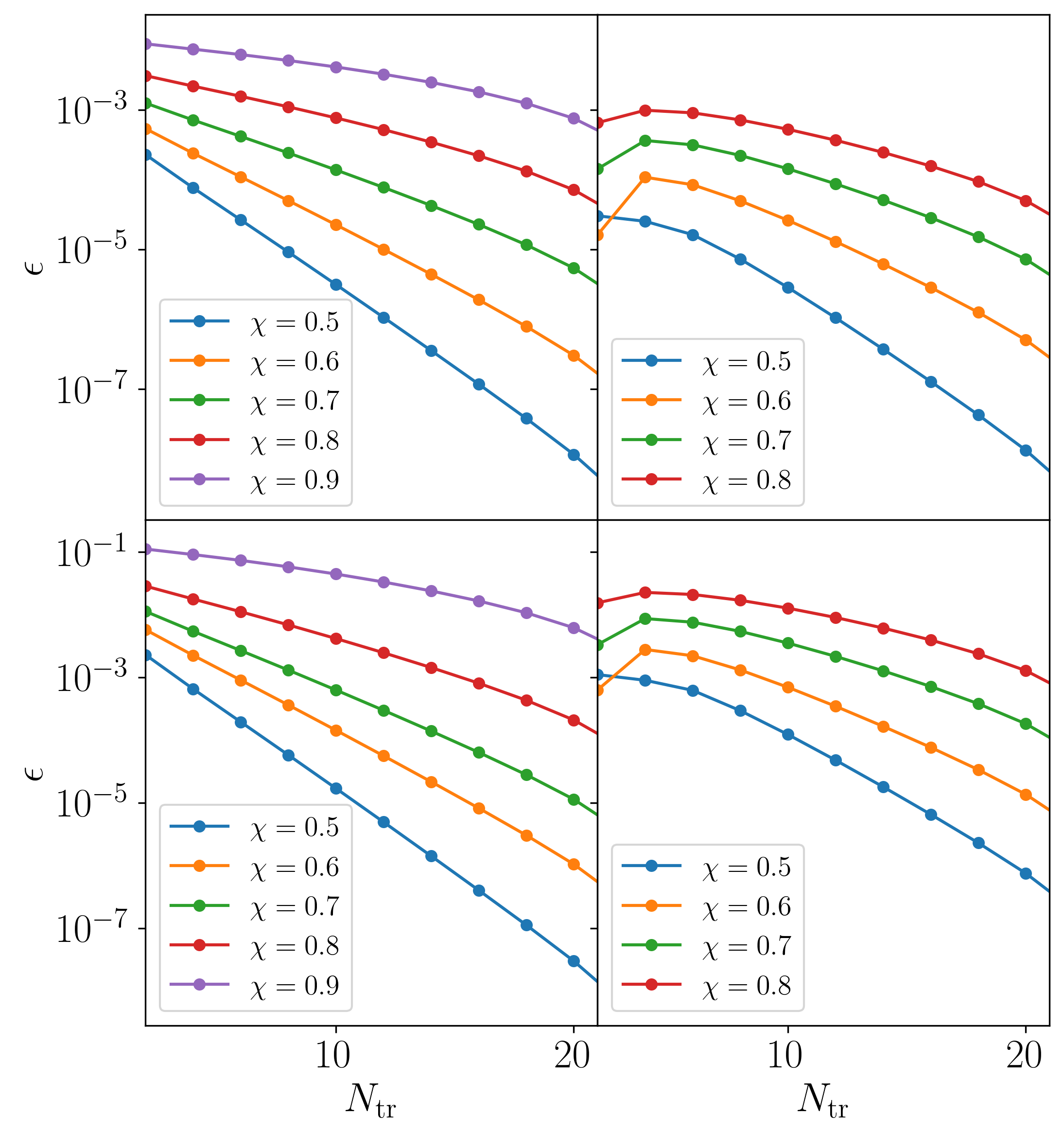}
    \includegraphics[scale=0.425]{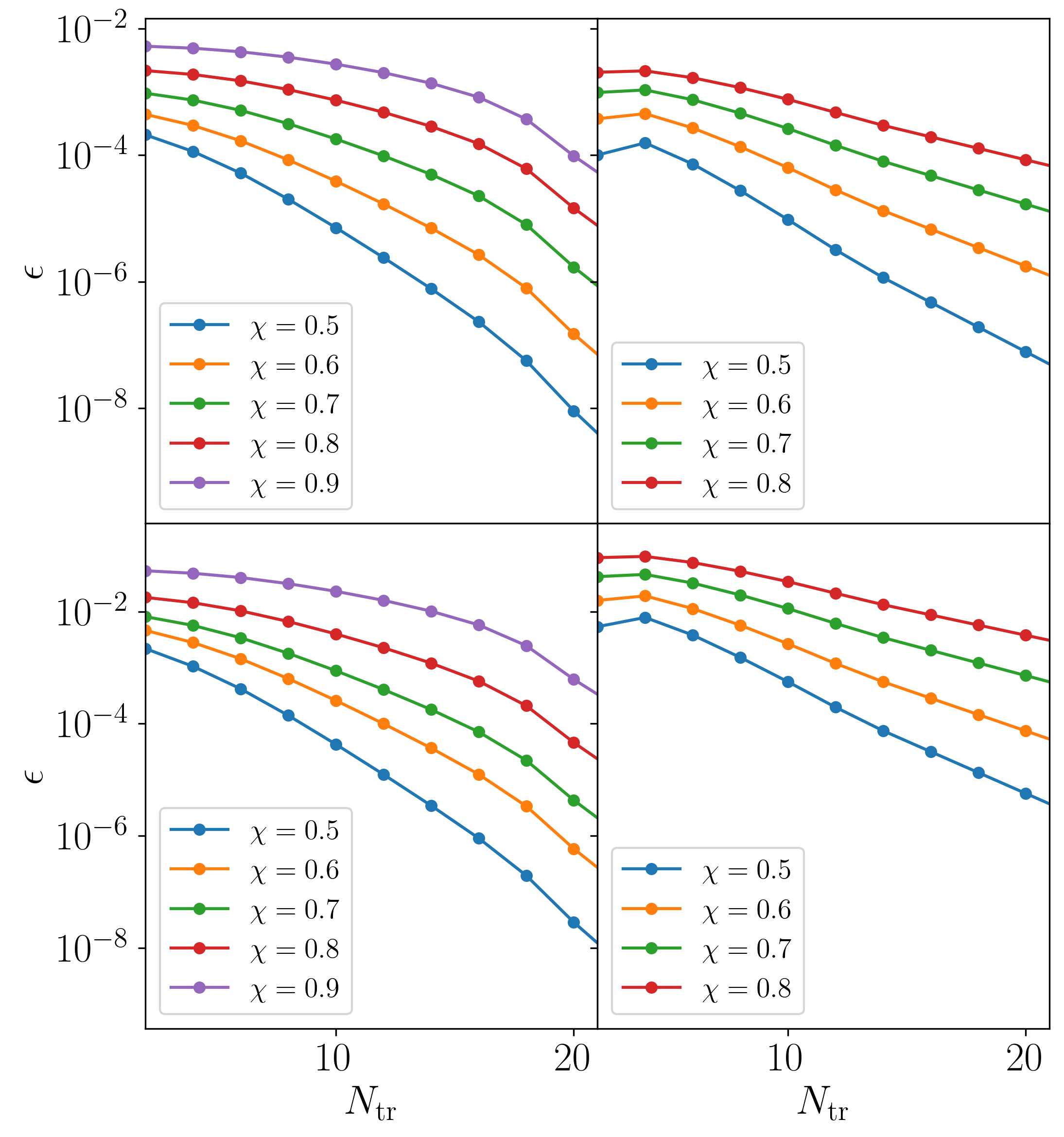}
    \caption{The accuracy of 4 observables as a function of $N_{\rm tr}$. Clockwise from top left: $M_2$, $R_{\rm ph}$, $L_{\rm ph}$, $\omega_{\rm ph}$.  The left columns are for dCS, with sGB on the right. The top row has $\zeta_{\rm q} = 0.1$, and the bottom row has $\zeta_{\rm q} = \zeta_{\rm max.}$.}
    \label{fig:relerror1}
\end{figure*}

\begin{figure*}[p]
    \centering
    \includegraphics[scale=0.425]{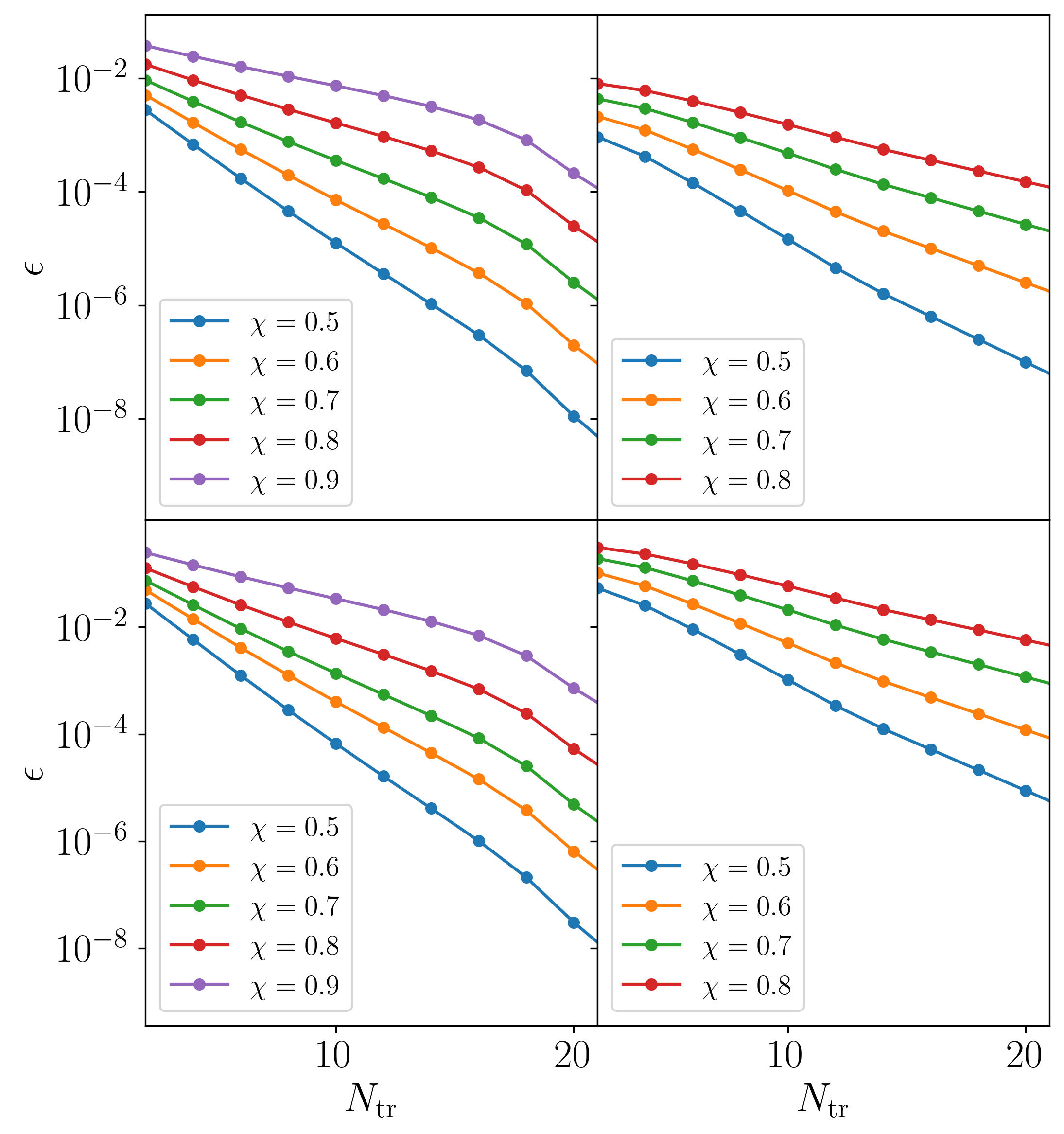}
    \includegraphics[scale=0.425]{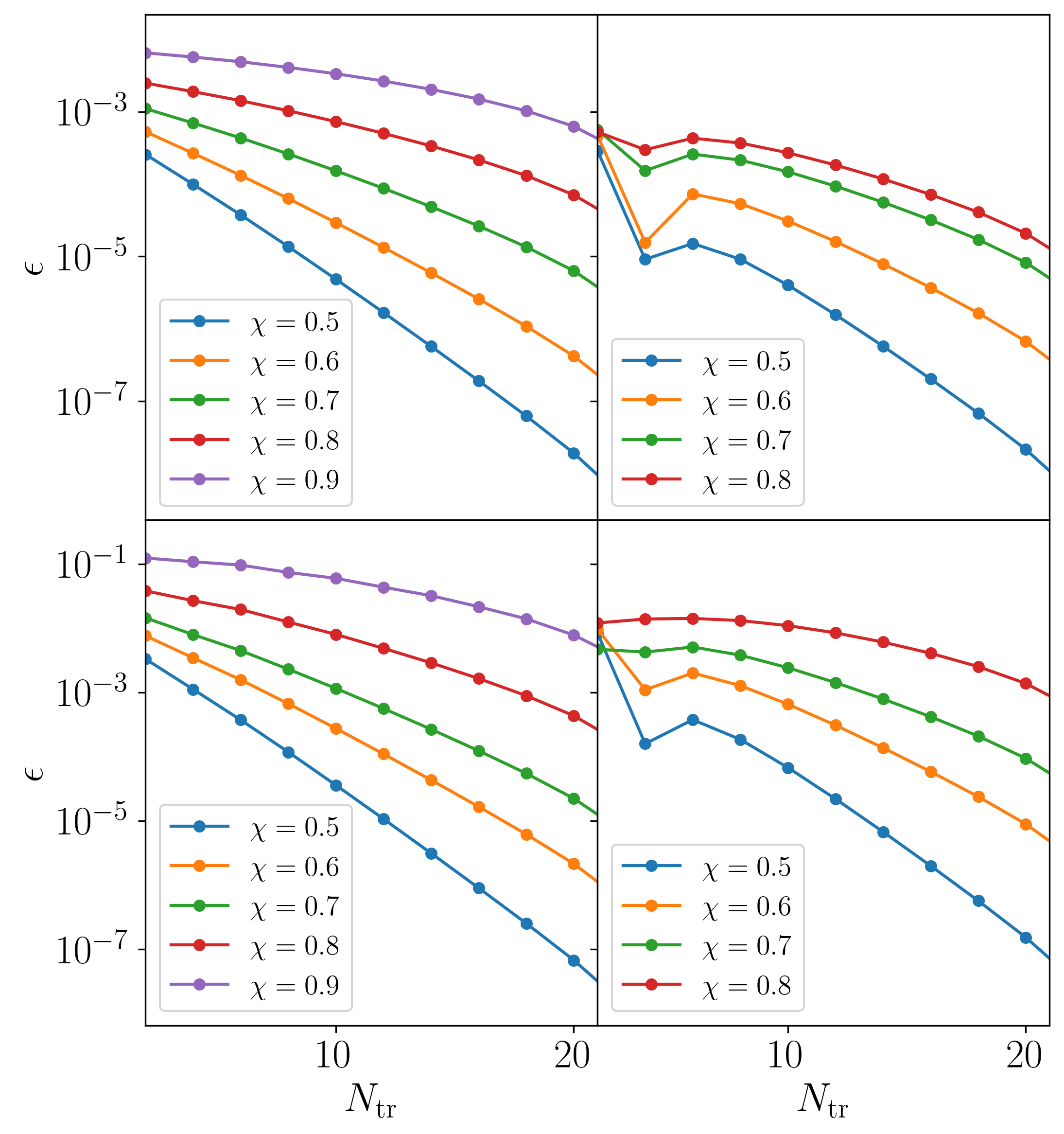} \\
    \includegraphics[scale=0.425]{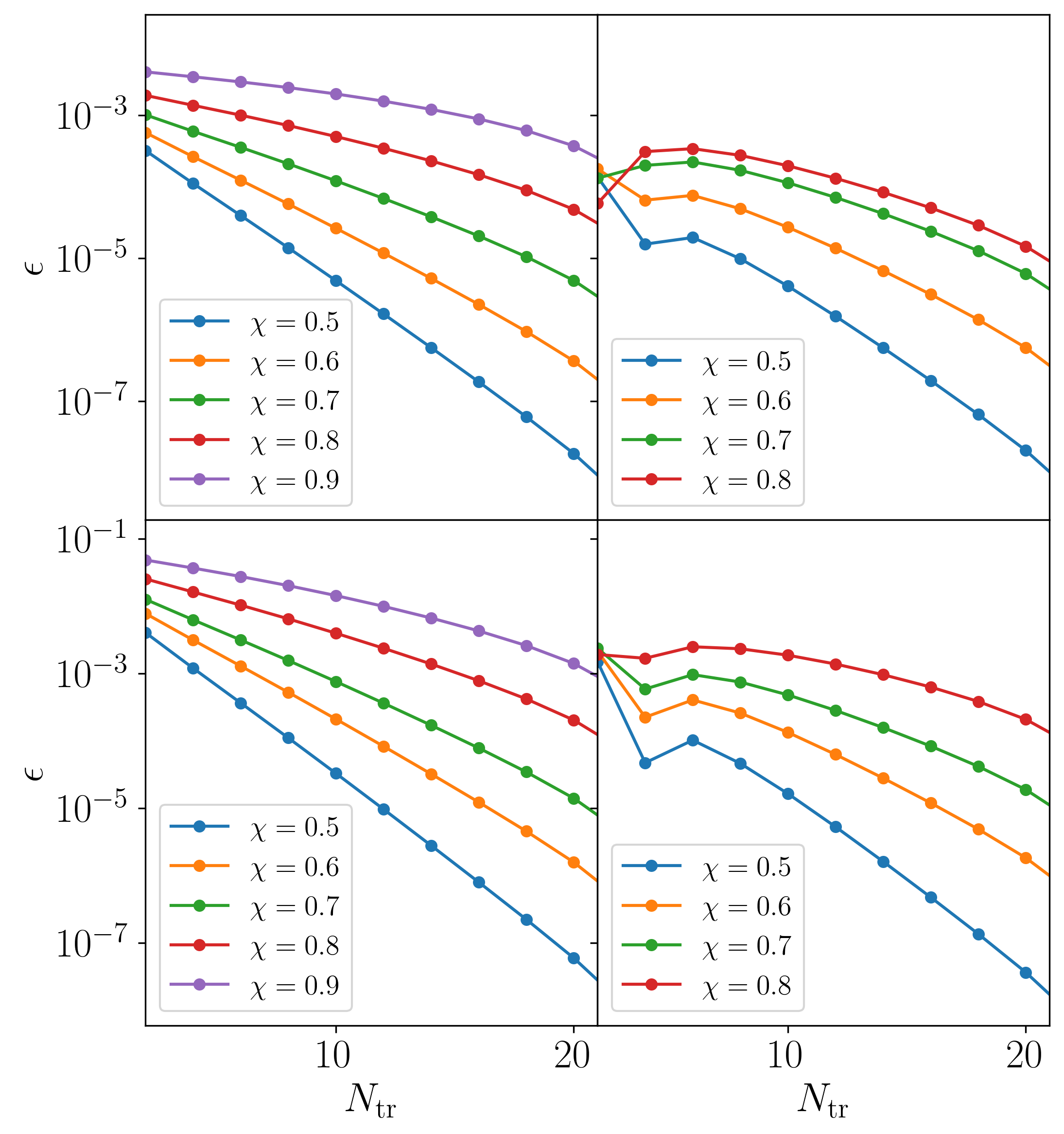}
    \includegraphics[scale=0.425]{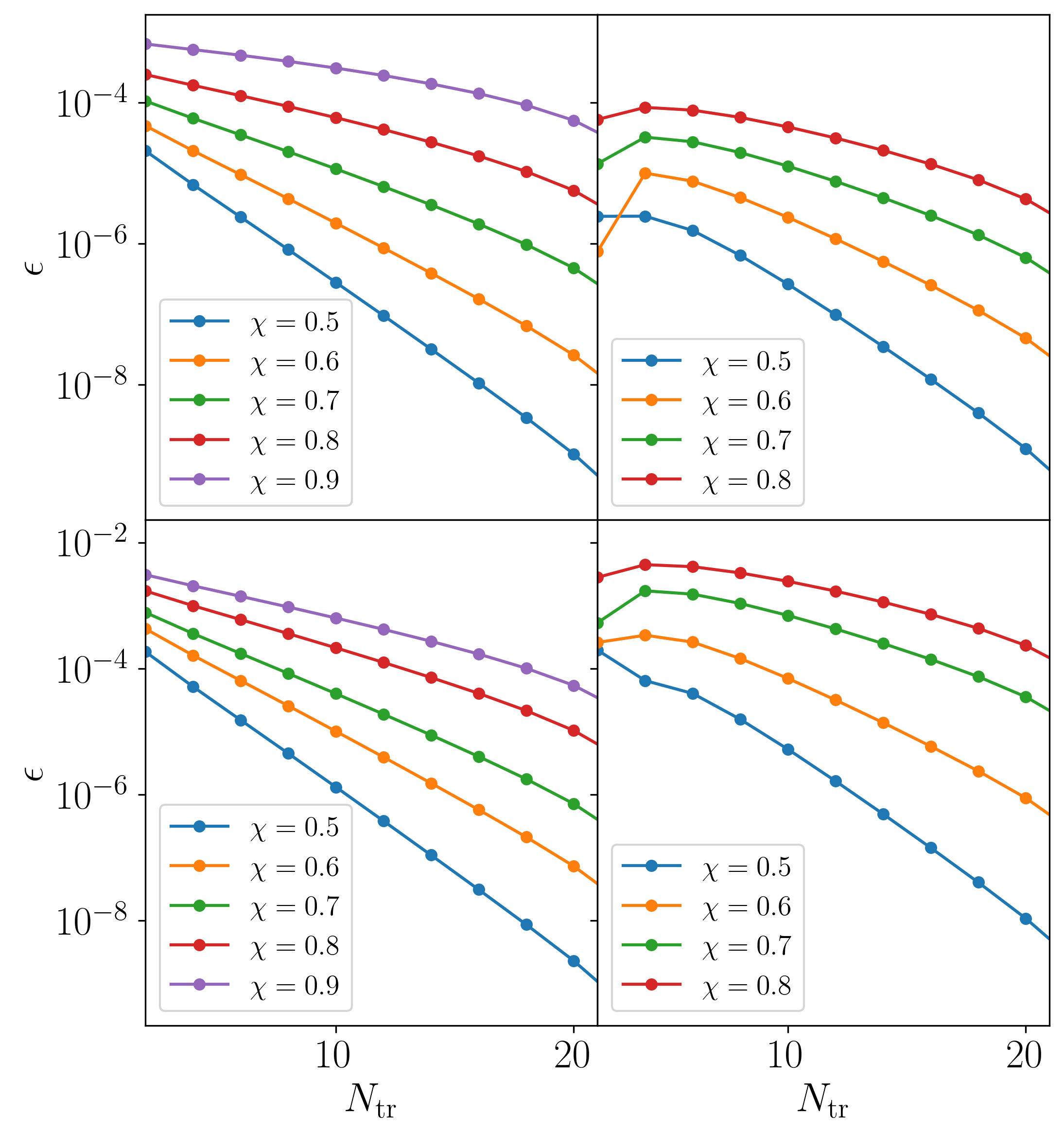}
    \caption{The accuracy of 4 observables as a function of $N_{\rm tr}$. Clockwise from top left: $\lambda_{\rm ph}$, $R_{\rm erg}$, $R_{\rm ISCO}$, $E_{b}$.  The left columns are for dCS, with sGB on the right. The top row has $\zeta_{\rm q} = 0.1$, and the bottom row has $\zeta_{\rm q} = \zeta_{\rm max.}$.}
\end{figure*}

\begin{figure*}[p]
    \centering
    \includegraphics[scale=0.425]{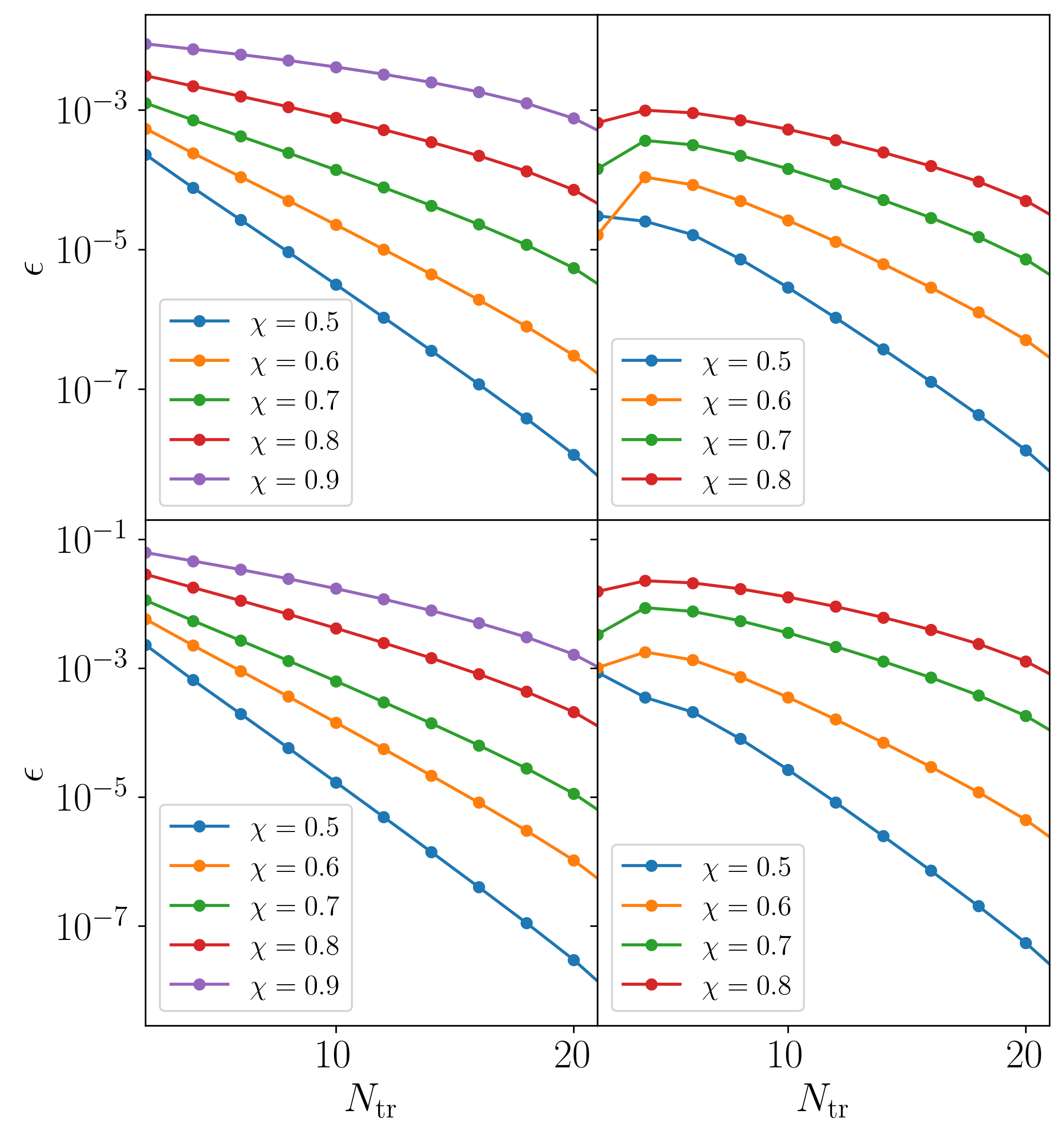}
    \includegraphics[scale=0.425]{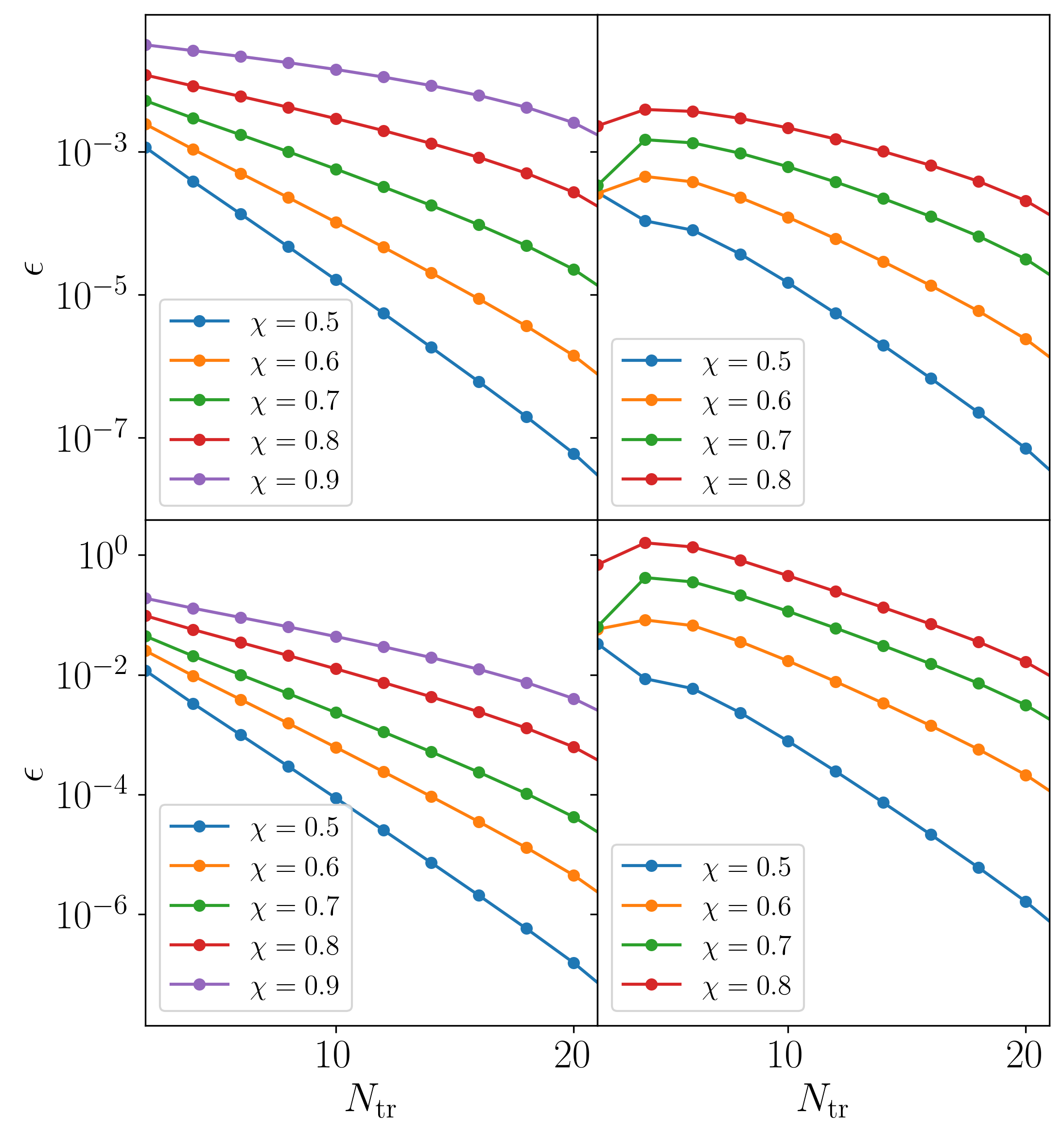}
    \caption{The accuracy of $L_{\rm ISCO}$ (left) and $\omega_{\rm ISCO}$ (right) as a function of $N_{\rm tr}$. The left columns are dCS, with sGB on the right. The top row has $\zeta_{\rm q} = 0.1$, and the bottom row has $\zeta_{\rm q} = \zeta_{\rm max.}$.}
    \label{fig:my_label}
\end{figure*}

\section{Required order plots}\label{sec:reqord}
Below we show the order required to achieve a given error for every observable listed in~\tab{tab:summary}.

\begin{figure*}[p]
    \centering
    \includegraphics[scale=0.425]{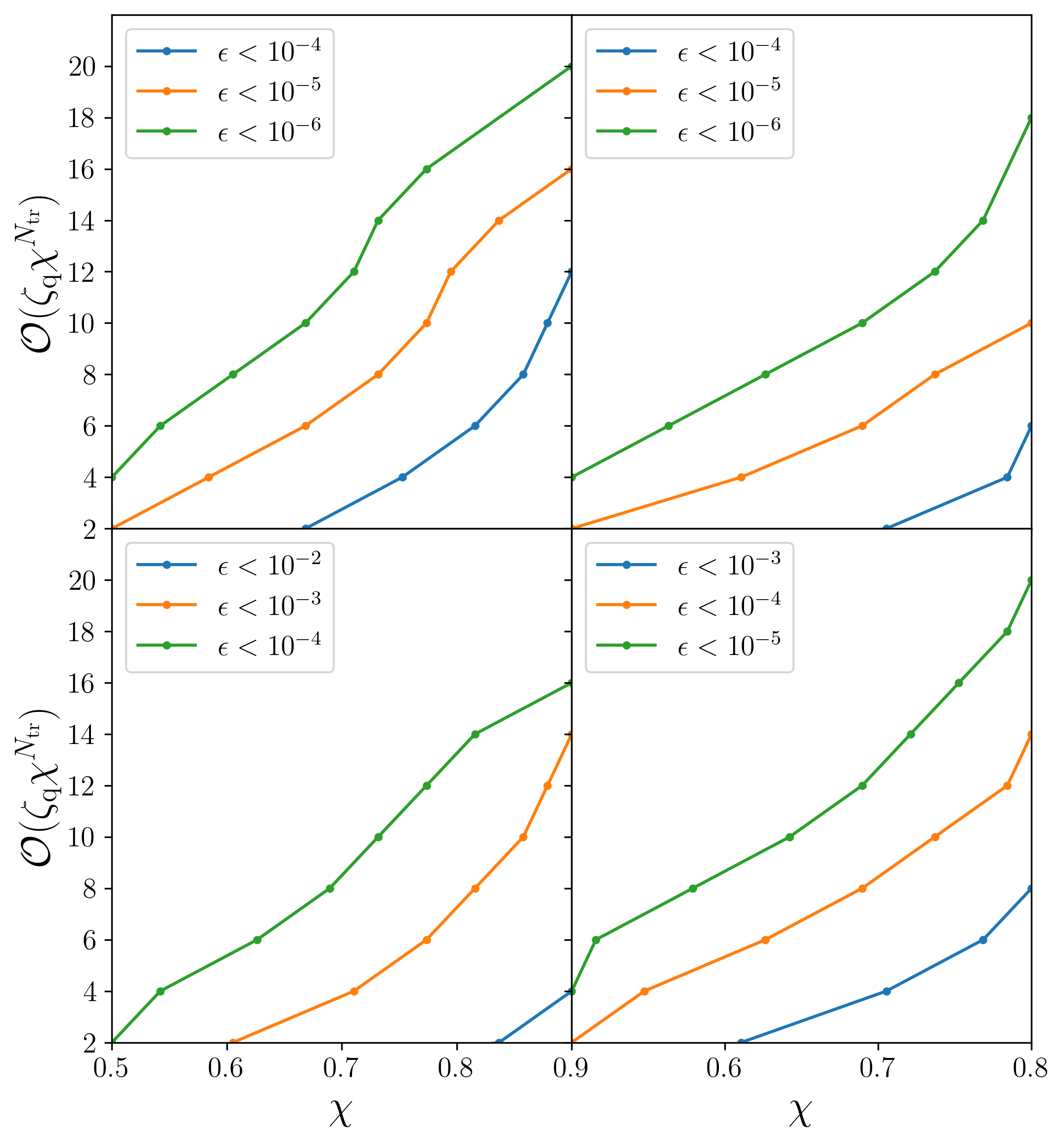}
    \includegraphics[scale=0.425]{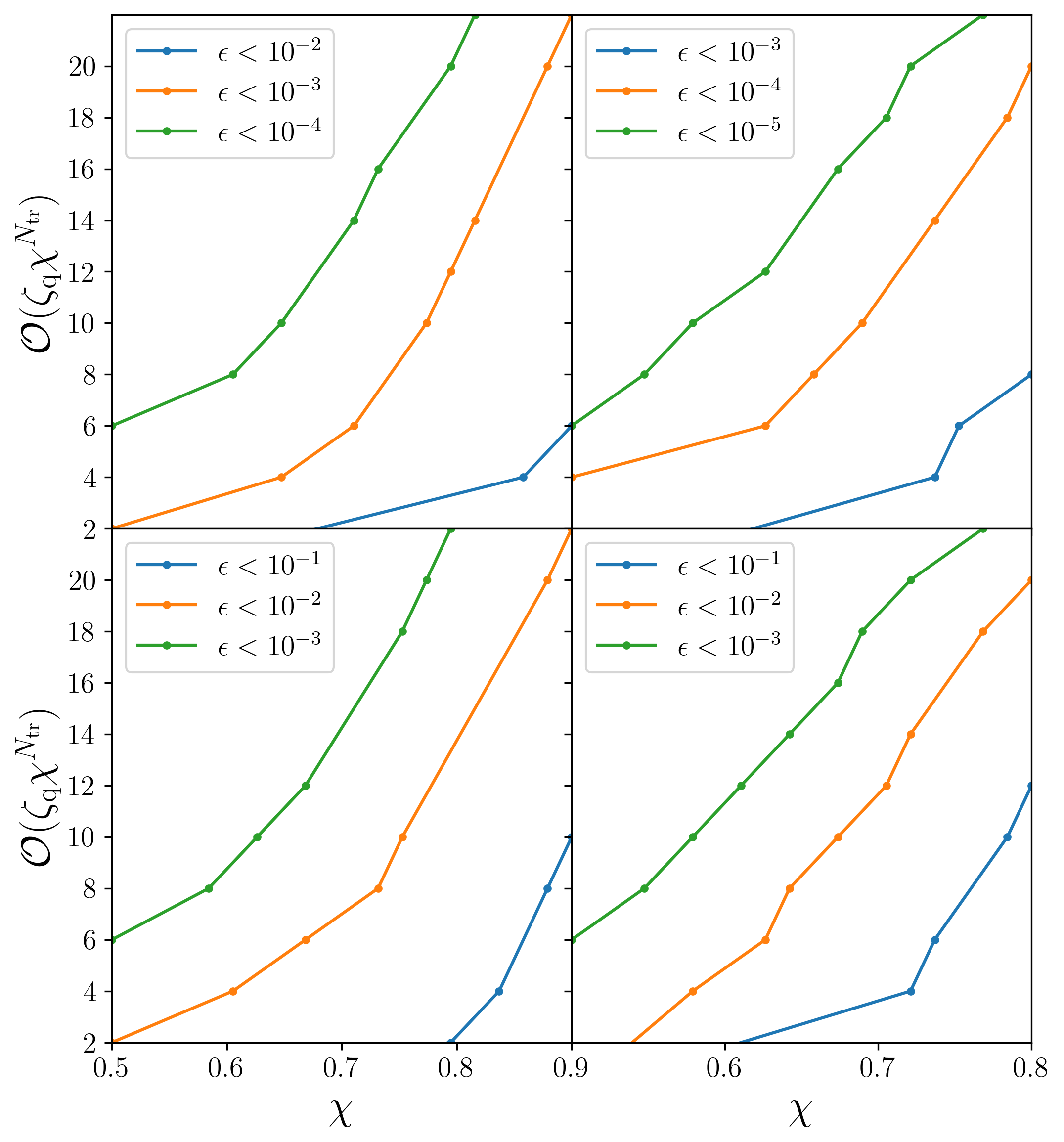} \\
    \includegraphics[scale=0.425]{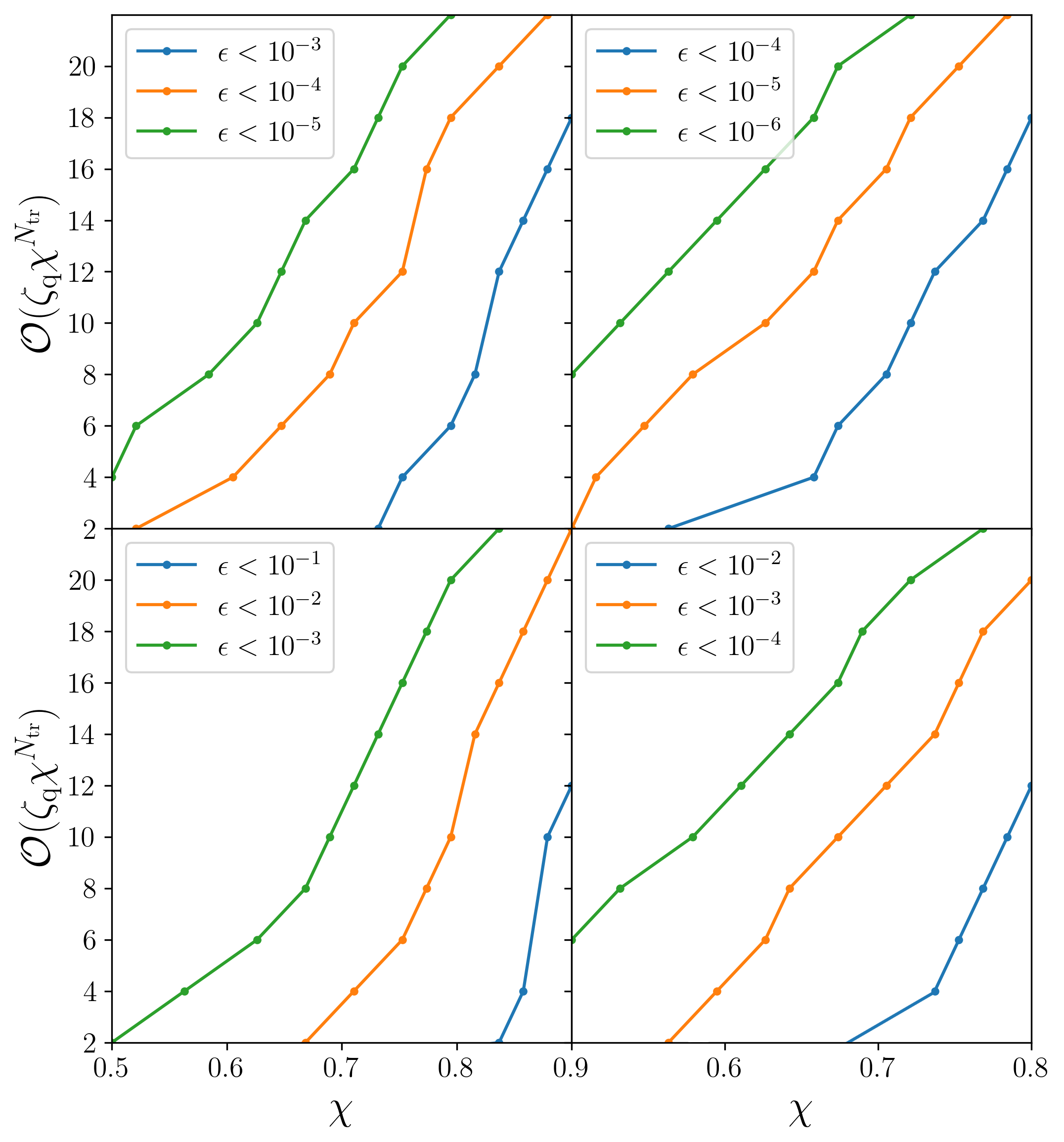}
    \includegraphics[scale=0.425]{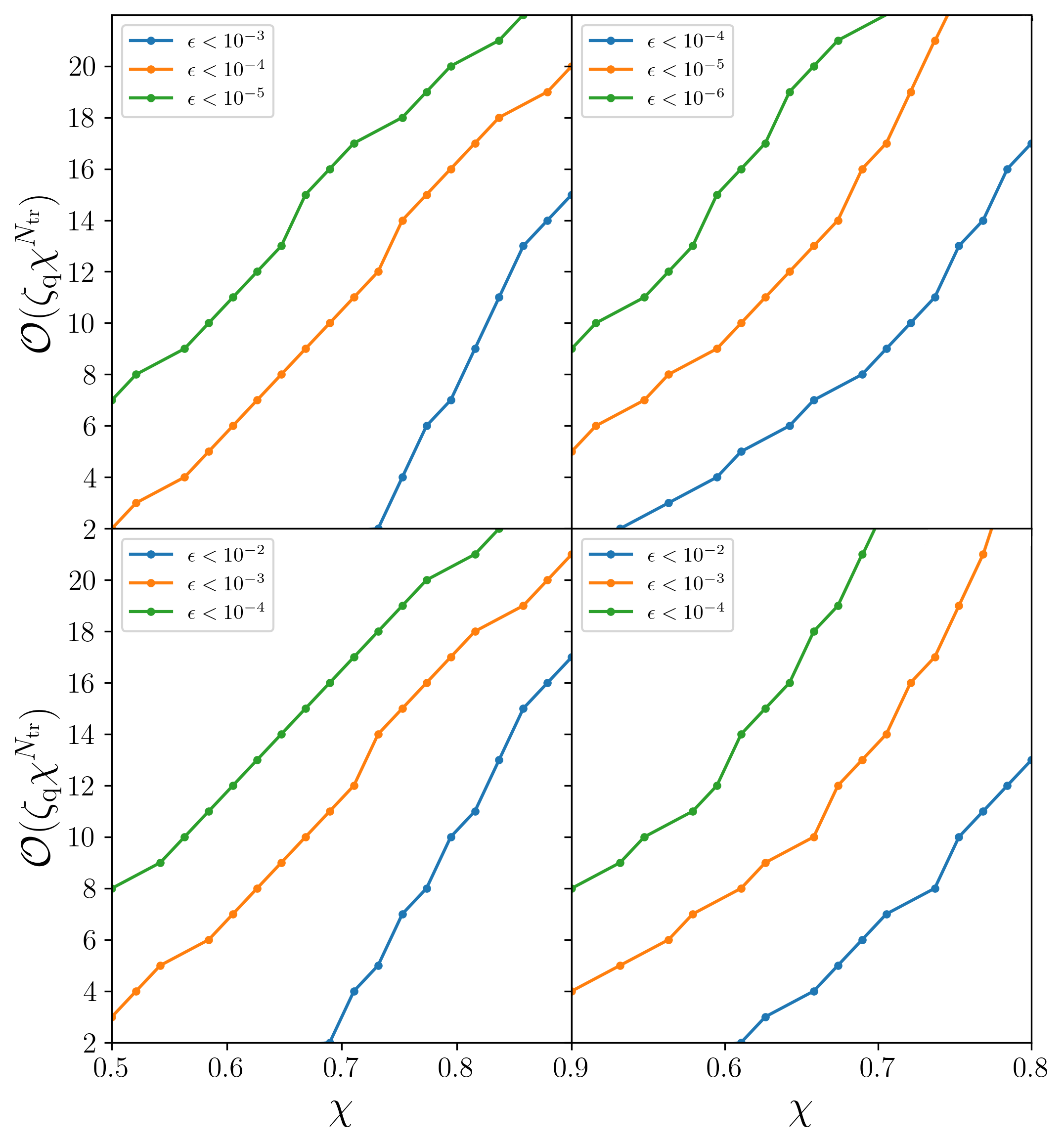}
    \caption{The required value of $N_{\rm tr}$ to achieve a given accuracy for 4 observables. Clockwise from top left: $M_2$, $R_{\rm ph}$, $L_{\rm ph}$, $\omega_{\rm ph}$.  The left columns are for dCS, with sGB on the right. The top row has $\zeta_{\rm q} = 0.1$, and the bottom row has $\zeta_{\rm q} = \zeta_{\rm max.}$.}
    \label{fig:requiredorder1}
\end{figure*}

\begin{figure*}[p]
    \centering
    \includegraphics[scale=0.425]{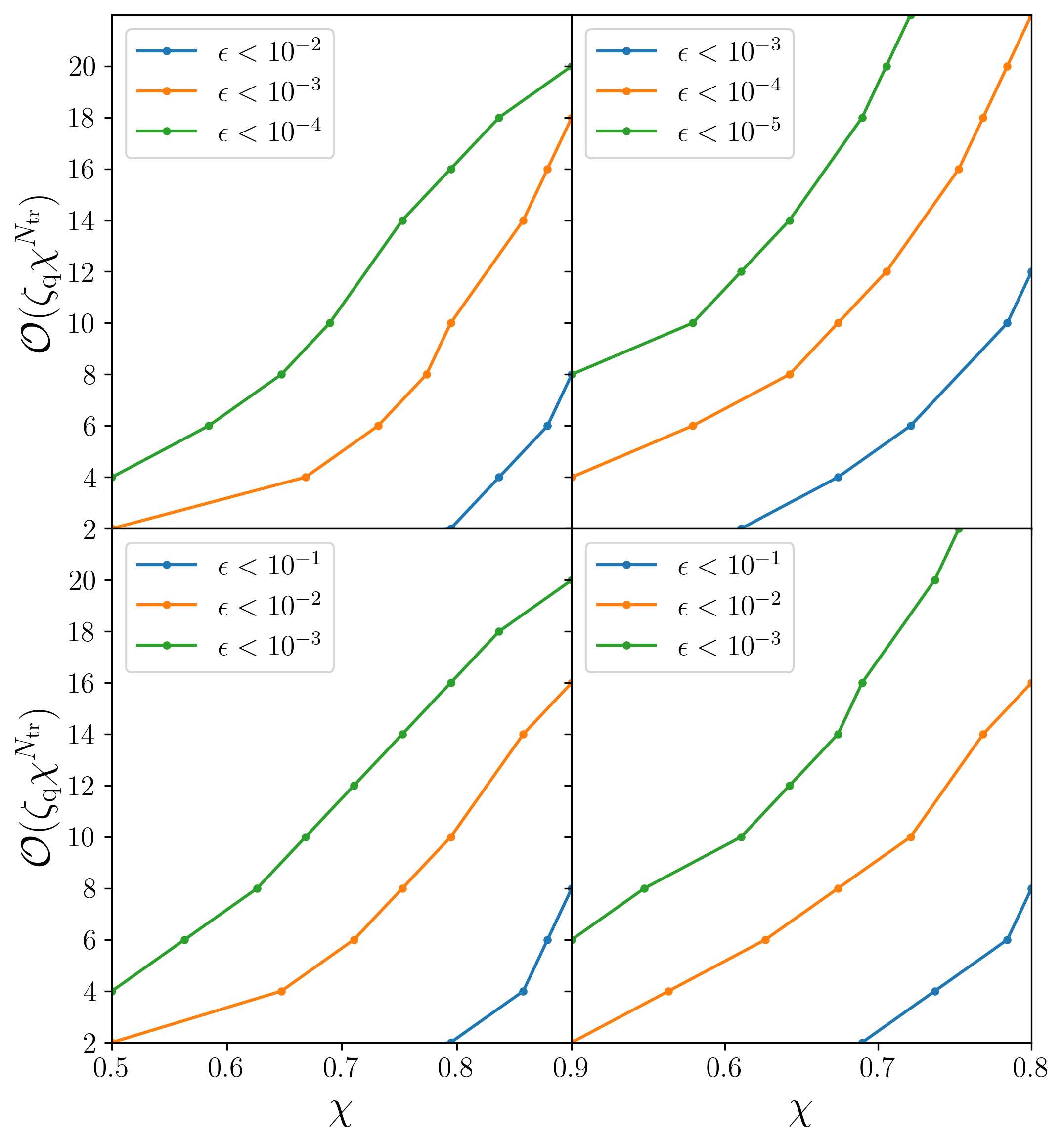}
    \includegraphics[scale=0.425]{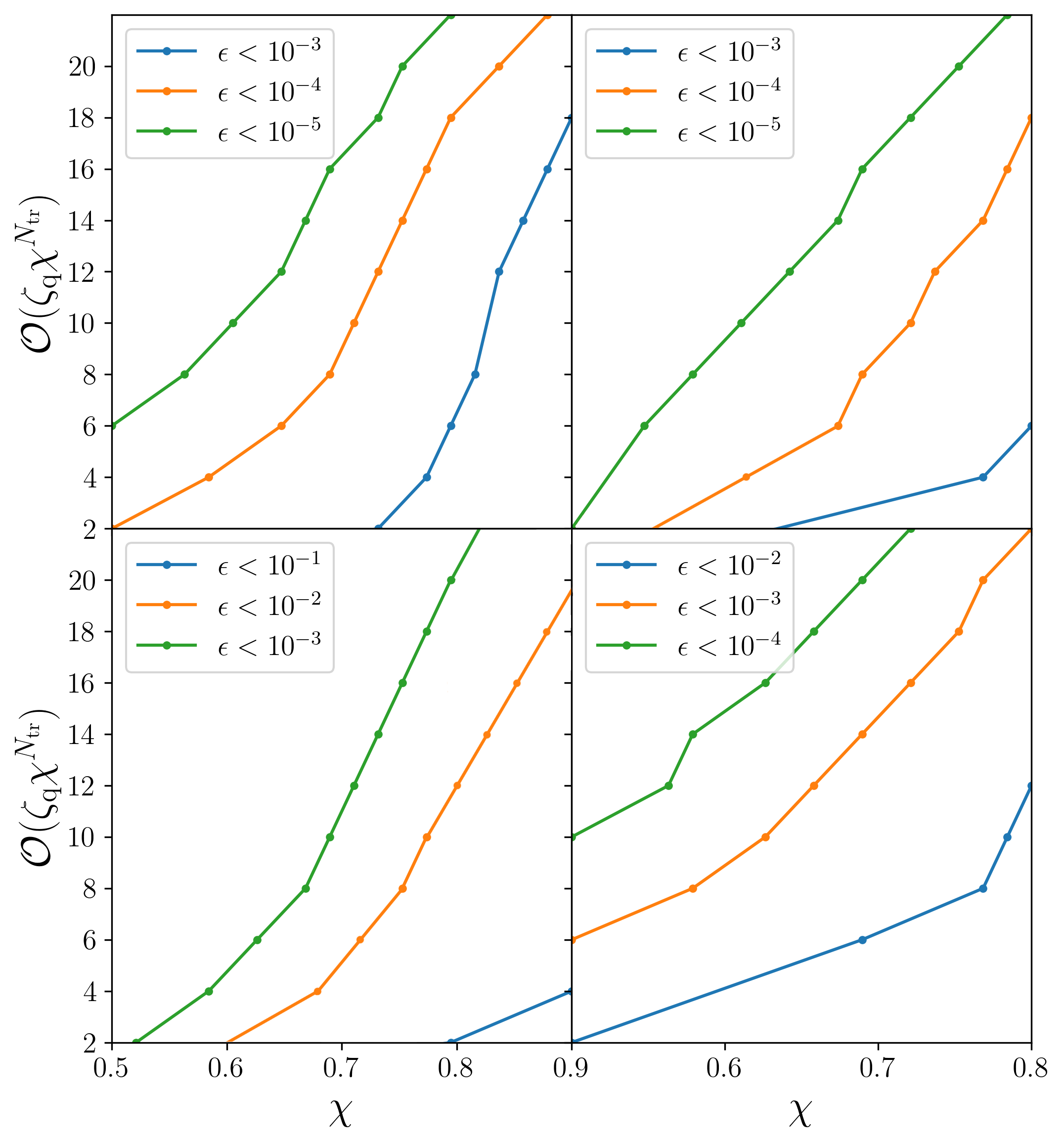} \\
    \includegraphics[scale=0.425]{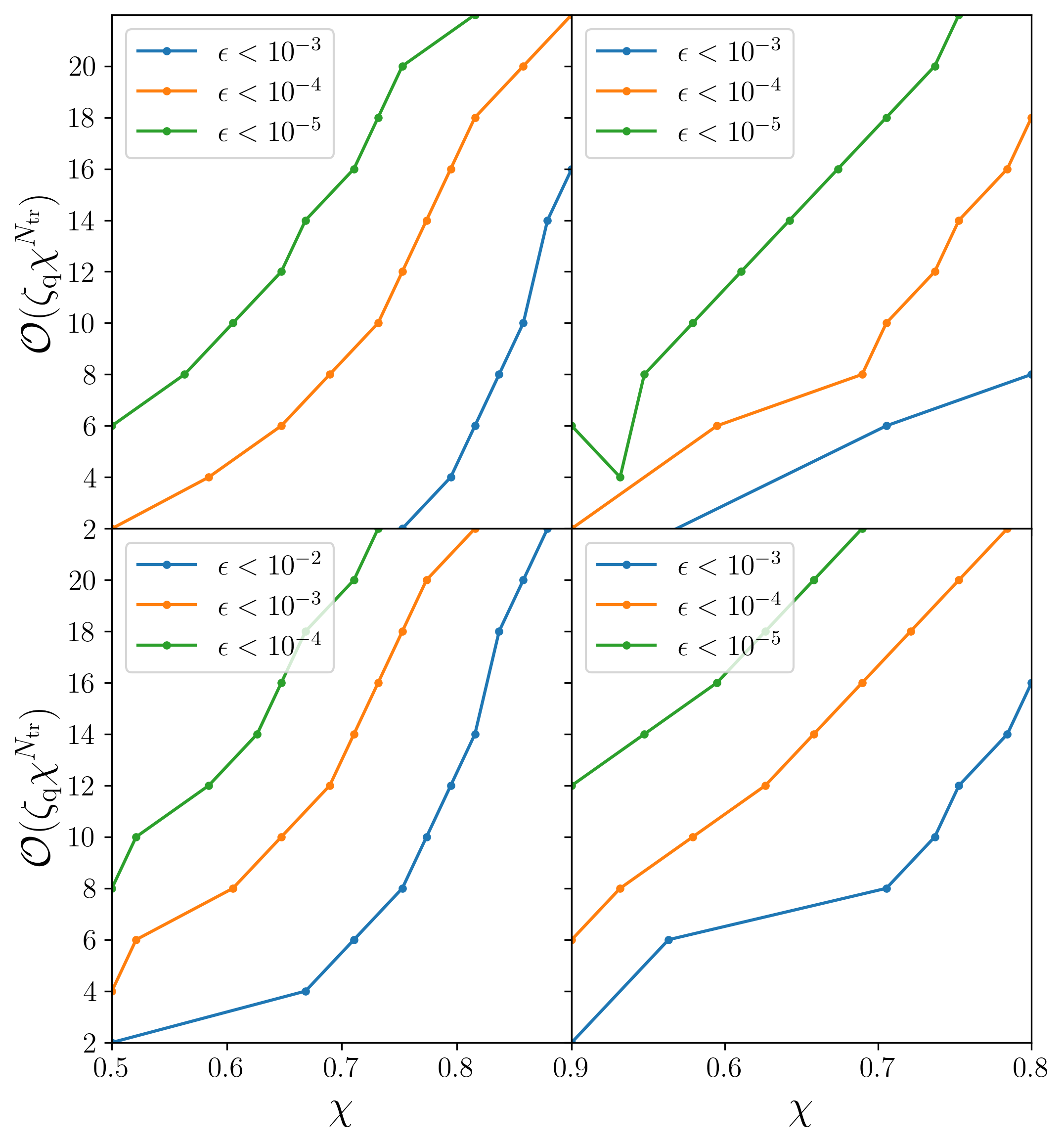}
    \includegraphics[scale=0.425]{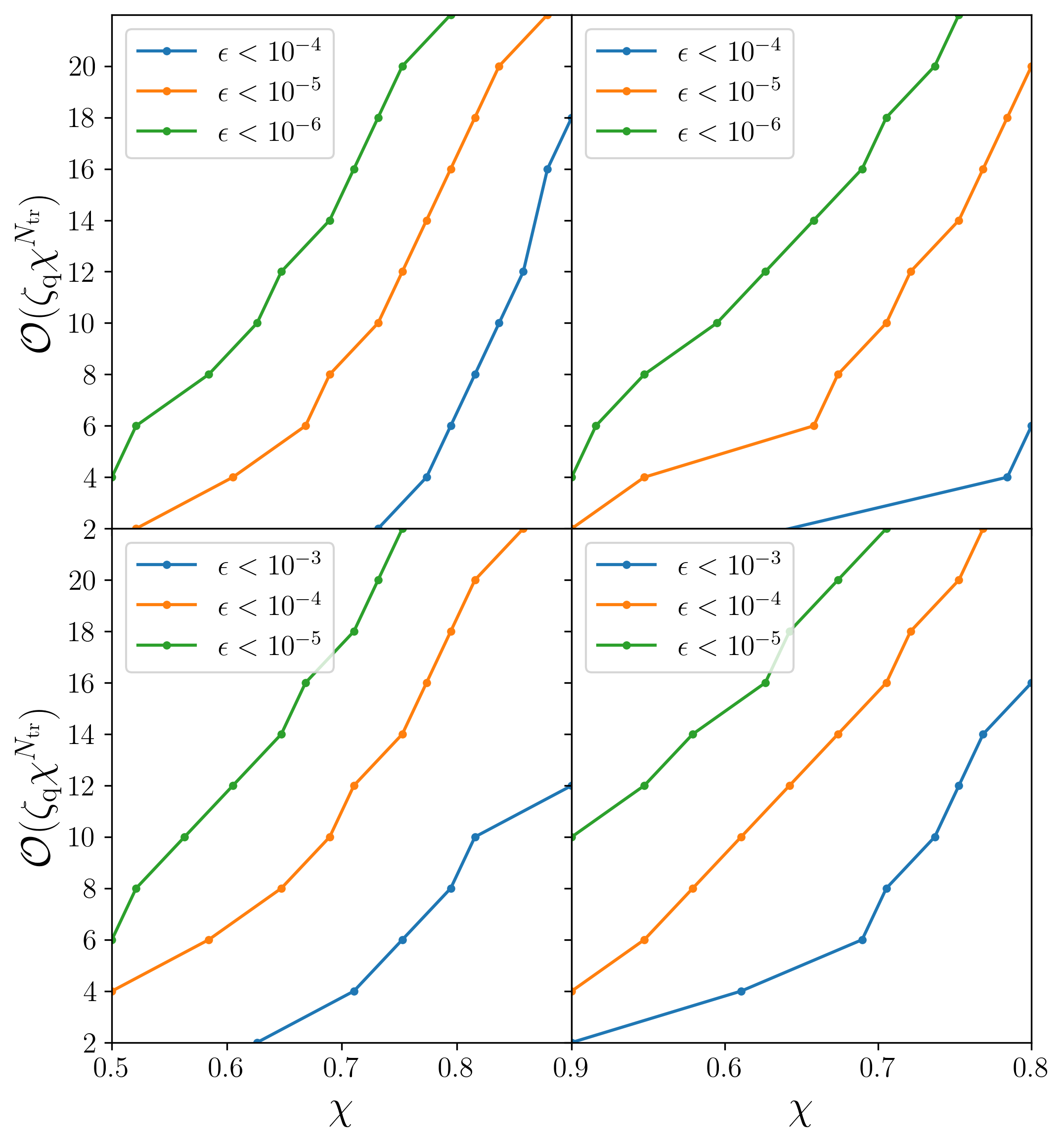}
    \caption{The required value of $N_{\rm tr}$ to achieve a given accuracy for 4 observables. Clockwise from top left: $\lambda_{\rm ph}$, $R_{\rm erg}$, $R_{\rm ISCO}$, $E_{b}$.  The left columns are for dCS, with sGB on the right. The top row has $\zeta_{\rm q} = 0.1$, and the bottom row has $\zeta_{\rm q} = \zeta_{\rm max.}$.}
    \label{fig:my_label}
\end{figure*}

\begin{figure*}[p]
    \centering
    \includegraphics[scale=0.425]{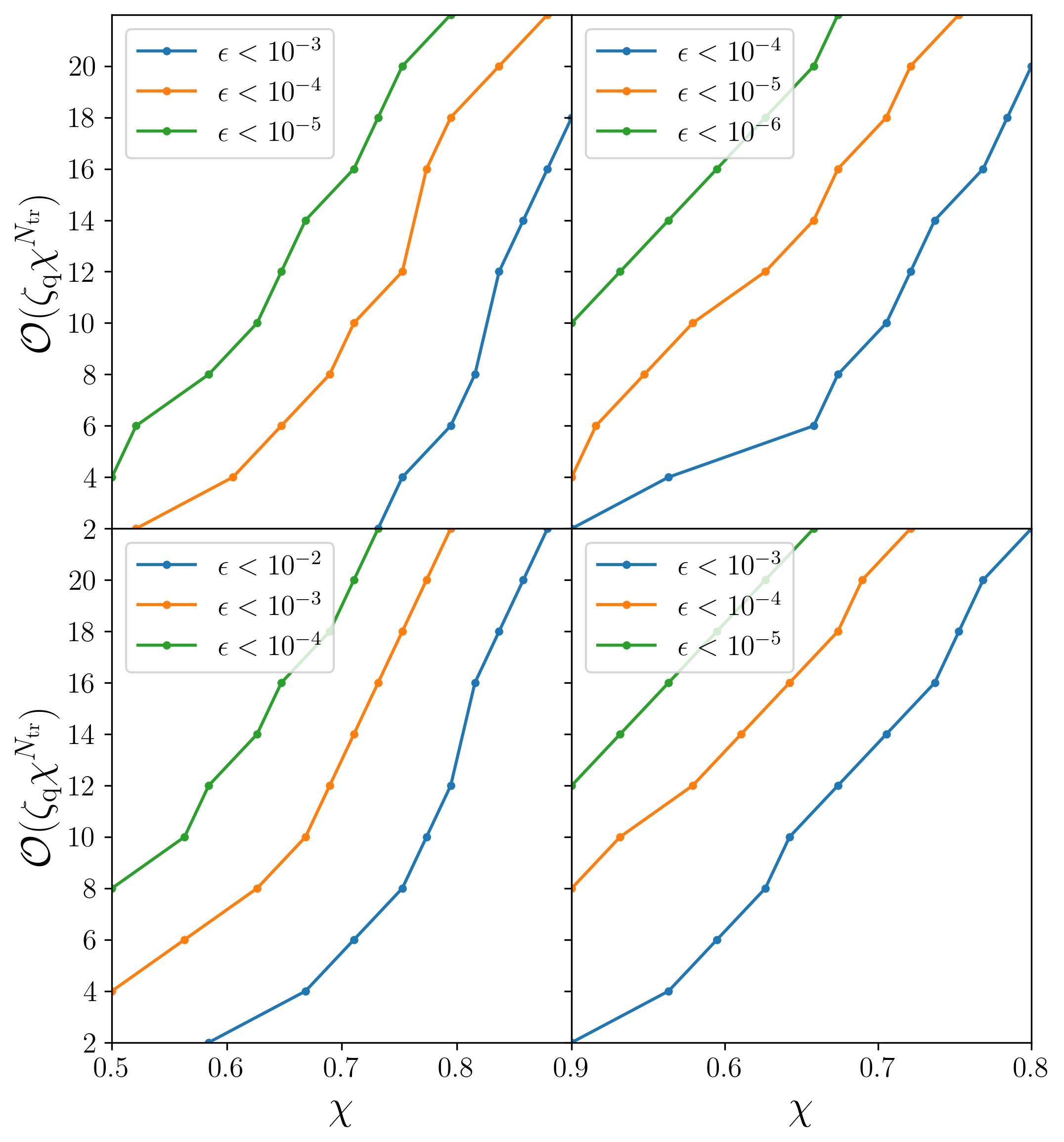}
    \includegraphics[scale=0.425]{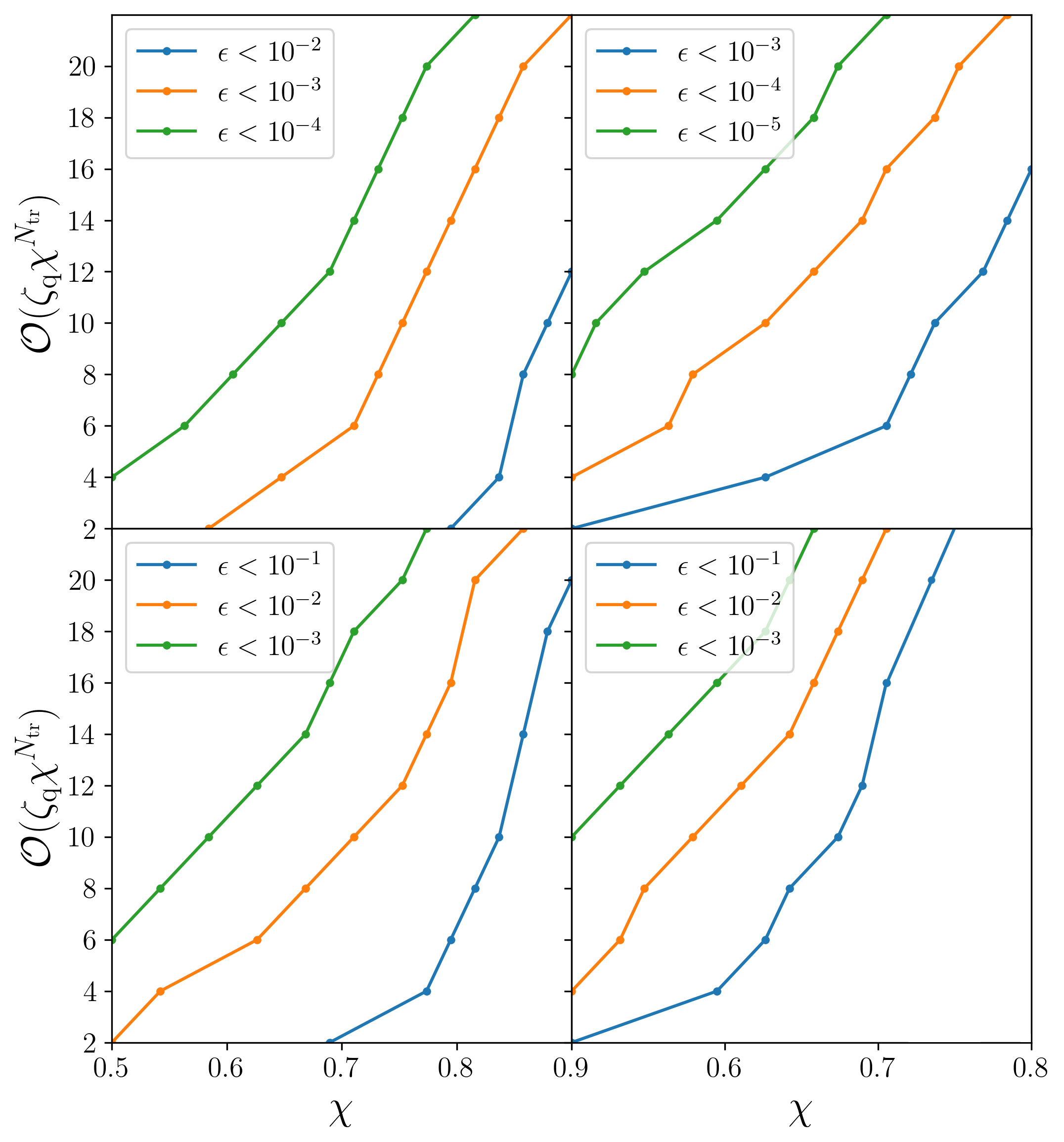}
    \caption{The required value of $N_{\rm tr}$ to achieve a given accuracy in $L_{\rm ISCO}$ (left) and $\omega_{\rm ISCO}$ (right). The left columns are dCS, with sGB on the right. The top row has $\zeta_{\rm q} = 0.1$, and the bottom row has $\zeta_{\rm q} = \zeta_{\rm max.}$.}
    \label{fig:my_label}
\end{figure*}

\bibliographystyle{apsrev4-1} 
\bibliography{tex/bib.bib}
\end{document}
%